\documentclass[aps, preprint, groupedaddress, superscriptaddress]{revtex4-1}

\pdfoutput=1

\usepackage{amsmath}
\usepackage{amssymb}
\allowdisplaybreaks[2]

\usepackage{graphicx}
\usepackage{subfigure}
\usepackage{longtable} 

\usepackage{multirow}

\usepackage{dcolumn}
\usepackage{bm}
\usepackage{array}
\usepackage{color}

\begin{document}

\title{A local algorithm and its percolation analysis of bipartite $z$-matching problem}

\author{Jin-Hua Zhao}
\email{zhaojh@m.scnu.edu.cn}

\affiliation{
School of Data Science and Engineering,
South China Normal University,
Shanwei 516600, China}

\affiliation{
Guangdong Provincial Key Laboratory of Nuclear Science,
Institute of Quantum Matter,
South China Normal University,
Guangzhou 510006, China}

\affiliation{
Guangdong-Hong Kong Joint Laboratory of Quantum Matter,
Southern Nuclear Science Computing Center,
South China Normal University,
Guangzhou 510006, China}

\date{\today}

\begin{abstract}
A $z$-matching on a bipartite graph is a set of edges,
among which each vertex of two types of the graph is adjacent to at most $1$ and at most $z$ ($\geqslant 1$) edges, respectively.
The $z$-matching problem concerns finding $z$-matchings with the maximum size.
Our approach to this combinatorial optimization problem is twofold.
From an algorithmic perspective,
we adopt a local algorithm as a linear approximate solver to find $z$-matchings on any graph instance,
whose basic component is a generalized greedy leaf removal procedure in graph theory.
From a theoretical perspective, on uncorrelated random bipartite graphs,
we develop a mean-field theory for percolation phenomenon underlying the local algorithm,
leading to an analytical estimation of $z$-matching sizes on random graphs.
Our analytical theory corrects the prediction by belief propagation algorithm at zero-temperature limit
in (Krea\v{c}i\'{c} and Bianconi 2019 \textsl{EPL} \textbf{126} 028001).
Besides, our theoretical framework extends a core percolation analysis of $k$-XORSAT problems to a general context of uncorrelated random hypergraphs with arbitrary degree distributions of factor and variable nodes.
\end{abstract}

\maketitle


\tableofcontents

\section{Introduction}

The analysis of algorithms and the computational complexity of combinatorial optimization problems
\cite{
Papadimitriou.Steiglitz-1998,
Pardalos.Du.Graham-2013-2e,
Garey.Johnson-1979}
on graphs and networks
\cite{
Bollobas-2002,Newman-2018}
are among the major interdisciplinary topics in applied mathematics, computer science, and statistical physics.
Once we map solving combinatorial optimization problems to finding ground states of corresponding physical systems with discrete states, statistical mechanics of disordered spin glassy systems
\cite{
Mezard.Parisi.Virasoro-1987,
Nishimori-2001,
Mezard.Montanari-2009}
offers principled mean-field frameworks, such as the replica trick and the cavity method. These frameworks work at finite and zero temperatures, and can both characterize the organization of solution space on graph ensembles and find approximate solutions on graph instances.
As a special form of mean-field theories, a percolation analysis of graphical combinatorial optimization problems has a geometric origin and is usually based on local optimal steps to construct a solution configuration.
This approach works directly at zero temperature and involves solving a percolation model underlying those local steps \cite{Stauffer.Aharony-1994},
whose transition phenomena signify certain structural transitions in solution space of optimization problems.
With a rather simple intuitive and in a concise analytical form, this approach provides an alternative perspective to ground states of optimization problems.

A typical example of the above local optimal steps of combinatorial optimization problems,
which finally lead to a percolation analysis, is the greedy leaf removal (GLR) procedure
\cite{
Karp.Sipser-IEEFoCS-1981,
Aronson.Frieze.Pittel-RandStrucAlgo-1998}.
This procedure is an iterative pruning process on graphs,
in which any vertex with one nearest neighbor (a leaf),
along with its sole neighbor (a root), is removed.
It results in core percolation on graphs
\cite{
Bauer.Golinelli-EPJB-2001,
Liu.Csoka.Zhou.Posfai-PRL-2012}.
A well-known optimization problem dealt with the GLR procedure is the maximum matching (MM) problem
\cite{
Lovasz.Plummer-1986}.
The MM problem can be defined on an undirected graph,
and it aims to find sets of edges sharing no common vertex (matchings) with the maximum size.
Results from belief propagation algorithms at zero-temperature limit extrapolated from finite-temperature case
\cite{
Zhou.OuYang-arXiv-2003,
Zdeborova.Mezard-JStatPhys-2006}
and an extended core percolation analysis based on the GLR procedure
\cite{Zhao.Zhou-JStatMech-2019}
show similar result on matching sizes,
yet the latter method follows a much more direct and simpler derivation process.
Further, the MM problem on a directed graph can be mapped onto its bipartite graph representation
\cite{
Hopcroft.Karp-SIAMJComput-1973}.
There are results from the cavity method at zero-temperature limit
\cite{
Liu.Slotine.Barabasi-Nature-2011},
yet a core percolation analysis
\cite{Zhao.Zhou-PRE-2019}
gives a corrected estimation of matching sizes in percolated regime.
The GLR procedure is also applied on the minimum vertex cover problem
\cite{
Bauer.Golinelli-EPJB-2001,
Zhao.Zhou-JStatMech-2019},
and reproduces in a simple way the results of energy densities from a calculation with replica trick on Erd\"os-R\'enyi (ER) random graphs
\cite{Weigt.Hartmann-PRL-2000}.
Variants of the GLR procedure are also adopted in $k$-XORSAT problem
\cite{
RicciTersenghi.Weigt.Zecchina-PRE-2001,
Franz.etal-PRL-2001,
Mezard.RicciTersenghi.Zecchina-JStatPhys-2003,
Cocco.etal-PRL-2003},
Boolean networks
\cite{Correale.etal-PRL-2006},
maximum independent set problem
\cite{Lucibello.RicciTersenghi-IntJStatMech-2014},
minimum dominating set problem
\cite{
Zhao.Zhou-JStatPhys-2015,
Zhao.Zhou-LNCS-2015},
and covering problems on hypergraphs
\cite{Coutinho.etal-PRL-2020}.
Besides its original definition based on leaves (any node with a degree of $1$) on undirected graphs,
the GLR procedure can further be generalized on directed graphs
\cite{AzimiTafreshi.Dorogovtsev.Mendes-PRE-2013}
and can also be based on $k$-leaves (any node with a degree $< k$ with $k$ as an integer) on both single and multiplex networks
\cite{AzimiTafreshi.Osat.Dorogovtsev-PRE-2019,
Shang-NJP-2019,
Shang-PRE-2020,
Shang-AdvInComp-2020}.

In this paper, we study the $z$-matching problem on bipartite graphs
\cite{
Gu.etal-IEEECommuMag-2015,
Kreacic.Bianconi-EPL-2019,
Kahlke.etal-EPJB-2021}.
This problem has a direct implication in resource allocation and sharing in various contexts due to
the complementary roles between different types of agents (for example, providers and consumers) in them
\cite{
Gu.etal-IEEECommuMag-2015}.
A $z$-matching ($z \geqslant 1$) is a set of edges, denoted as matched edges, of a bipartite graph,
such that any vertex of one type is adjacent to at most $1$ matched edge
and any vertex of the other type is adjacent to at most $z$ matched edges.
The $z$-matching problem concerns finding $z$-matchings with the maximum size.
Our main contributions in this paper are as follows.
(1) On the algorithmic side, we approach the $z$-matching problem with a local algorithm based on a generalized GLR procedure on bipartite graphs (GLRB), which is a linear approximation method.
(2) On the theoretical side, on uncorrelated random bipartite graphs,
we develop an analytical theory to elucidate the percolation phenomenon underlying GLRB,
and further derive an analytical estimation of $z$-matching sizes.
(3) Compared with a previous mean-field theory extrapolating from finite to zero temperature in \cite{Kreacic.Bianconi-EPL-2019}, our analytical framework is established directly at zero temperature, and provides a consistent result on ground state energy in a percolated phase.
(4) Our model and theory naturally extend a core percolation analysis of the $k$-XORSAT problem from the original context of each constraint with a fixed size of variables to a general one with arbitrary uncorrelated degree distributions for factor and variable nodes.

The layout of the paper is as follows.
In section \ref{sec:model}, we explain the $z$-matching problem and our local algorithm.
In section \ref{sec:theory}, we develop an analytical theory for the percolation phenomenon underlying the local algorithm, and further estimate $z$-matching sizes. We also compare our analytical framework with a previous mean-field theory. Then we apply our framework on $k$-XORSAT problem with constraints composing varying sizes of variables.
In section \ref{sec:results}, we test our model and theory on some random bipartite graph models and a real-world network.
In section \ref{sec:conclusion}, we conclude the paper.

\section{Model}
\label{sec:model}

%
\begin{figure}
\begin{center}
 \includegraphics[width = 0.99 \linewidth]{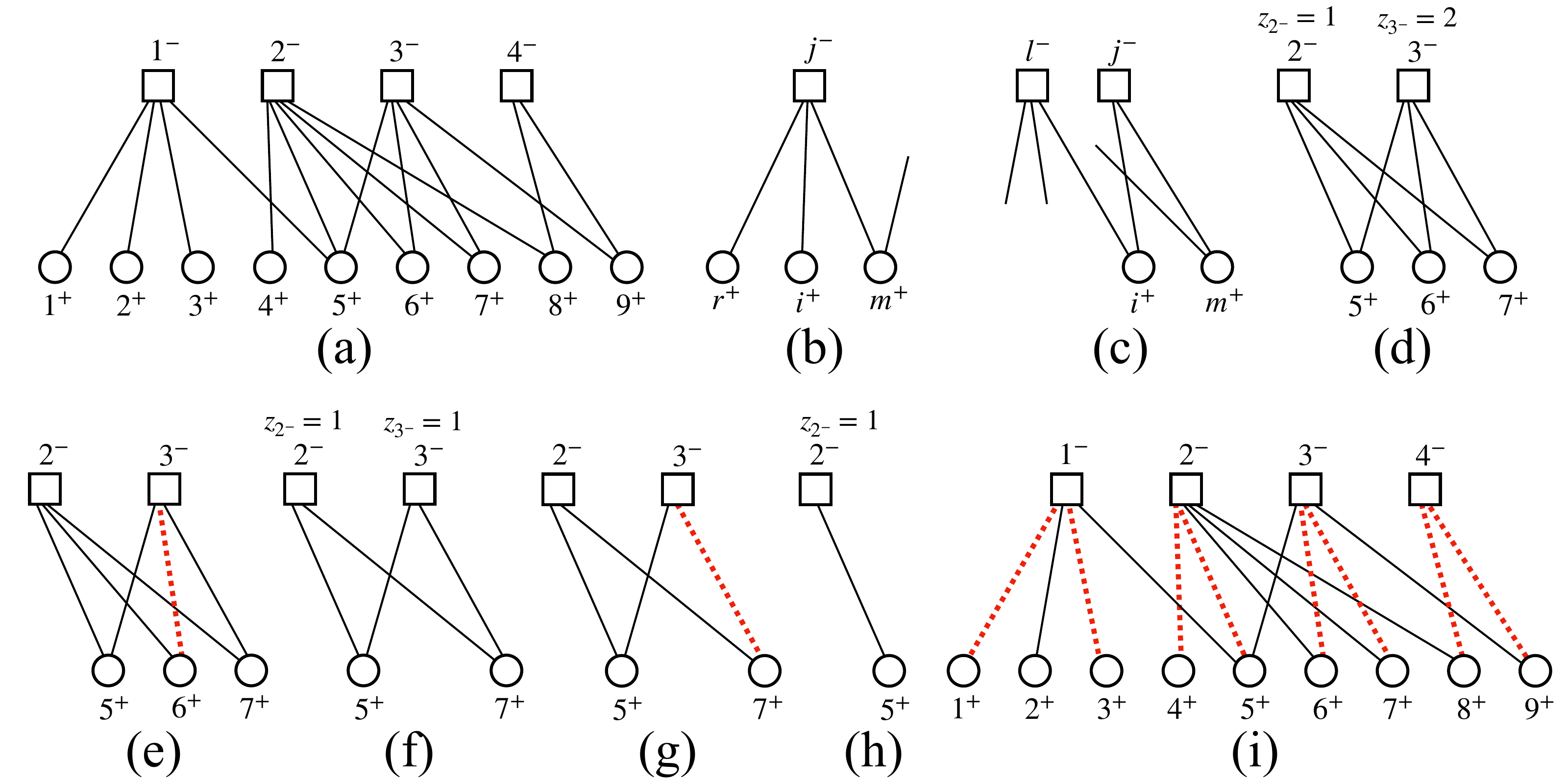} 
\end{center}
\caption{
\label{fig:model}
A local algorithm for the $z$-matching problem on a bipartite graph. Here we set $z = 2$.
(a) A bipartite graph with $9$ out-vertices (in circles), $4$ in-vertices (in squares), and $15$ edges among them.
(b) $r^{+}$ and $i^{+}$ are two out-leaves and $j^{-}$ is an in-root correspondingly. According to GLRB, edges $(r^{+}, j^{-})$ and $(i^{+}, j^{-})$ are matched and $z_{j^{-}}$ is updated to $0$, then all the edges adjacent to $j^{-}$ are further removed.
(c) $j^{-}$ is an in-leaf as $k_{j^{-}} \leqslant z_{j^{-}} (= 2)$, and $i^{+}$ and $m^{+}$ are out-roots correspondingly. According to GLRB, edges $(i^{+}, j^{-})$ and $(m^{+}, j^{-})$ are matched and $z_{j^{-}}$ is set to $0$, then all the edges adjacent to $i^{+}$, $m^{+}$, and $j^{-}$ are removed.
(d) The core of (a) after GLRB. Beware that following GLRB, as $4^{+}$ is an out-leaf, the edge $(4^{+}, 2^{-})$ is matched and $z_{2^{-}}$ is reduced by $1$.
(e) A local step is applied on the core, as an edge $(6^{+}, 3^{-})$ is chosen and matched.
(f) On (e), $z_{3^{-}}$ is reduced by $1$, and $6^{+}$ is removed. The current subgraph is also a core.
(g) On (f), a second local step is carried out, in which an edge $(7^{+}, 3^{-})$ is chosen and matched.
(h) On (g), $7^{+}$ is removed and $3^{-}$ is also removed as $z_{3^{-}} = 0$. Finally, $(5^{+}, 2^{-})$ is matched as $5^{+}$ is both an out-leaf and out-root.
(i) Taken the steps in (b)-(h) together, a $2$-matching solution with $8$ matched edges for (a) is denoted in red dashed lines.}
\end{figure}
%

First, we lay down some notations for bipartite graphs.
A bipartite graph $B = \{V_{+}, V_{-}, E\}$ consists of an out-vertex set $V_{+}$, an in-vertex set $V_{-}$, and an edge set $E = V_{+} \times V_{-}$ only between out- and in-vertices.
We define the size ratio of $B$ as $\phi \equiv |V_{-}| / |V_{+}|$. For any vertex $i^{+} \in V_{+}$, all its nearest neighbors in $V_{-}$ forms a set $\partial i^{+}$, and the degree of $i^{+}$ is $k_{i^{+}} \equiv |\partial i^{+}|$. Relevant notations can be defined on any vertex $j^{-} \in V_{-}$ for its neighbor set $\partial j^{-}$ and its degree $k_{j^{-}} \equiv |\partial j^{-}|$.

Factor graphs
\cite{Kschischang.Frey.Loeliger-IEEEInfTheor-2001}
are a special kind of bipartite graphs.
They are usually defined in a context involving complex functions, which can be factorized into local terms with a subset of variables. On the structural side, they are comprised of three basic elements: factor nodes being denoted as local functions, variable nodes as variables, and edges only connecting variable and factor nodes as a variable being included in a certain local function.
Factor graphs are the unified structural formalism for various subjects in disciplines ranging from computer science to information theory
\cite{Kschischang.Frey.Loeliger-IEEEInfTheor-2001},
and also provide a natural representation for many-body or higher-order interactions in networked systems
\cite{
Battiston.etal-PhysRep-2020,
Battiston.etal-NatPhys-2021}.
In the definition of the $z$-matching problem on bipartite graphs, nontrivial constraints for a proper matching configuration are mainly on in-vertices. Thus following the language for factor graphs, the out- and in-vertices in a bipartite graph here can easily be considered as variable and factor nodes, respectively.

For optimization problems, a local algorithm on a graph instance is both a protocol of selecting vertices or edges into a solution configuration and a graph pruning process.
In the $z$-matching problem,
for any in-vertex $j^{-} \in V^{-}$,
we define a parameter $z_{j^{-}}$ 
as the maximum number of matched edges to which $j^{-}$ can be adjacent.
Initially, $z_{j^{-}} = z$ for all $j^{-} \in V^{-}$.
After an edge $(i^{+}, j^{-}) \in E$ is matched, $z_{j^{-}}$ is reduced by $1$.
If $z_{j^{-}} = 0$, $j^{-}$ is removed along with all its adjacent edges.
As an out-vertex can only be adjacent to at most one matched edge,
$i^{+}$ is removed along with all its adjacent edges.

Our local algorithm for $z$-matching problem starts with GLRB,
which is an iterative removal of four local structures.
They are identified as:
(1) an out-leaf: an out-vertex with only one neighboring in-vertex;
(2) an in-root: an in-vertex, say $j^{-}$, having $\geqslant z_{j^{-}}$ neighboring out-leaves;
(3) an in-leaf: an in-vertex, say $j^{-}$, with a degree $k_{j^{-}} \leqslant z_{j^{-}}$;
(4) an out-root: an out-vertex neighboring an in-leaf.
See figures \ref{fig:model} (b) and (c) for an example.
GLRB goes as follows.
(1) For an out-leaf, say $i^{+}$, with $j^{-}$ as its sole neighboring in-vertex which is not an in-root,
the edge $(i^{+}, j^{-})$ is matched and $z_{j^{-}}$ is reduced by $1$;
correspondingly, $i^{+}$ is removed.
(2) For an in-root, say $j^{-}$,
$z_{j^{-}}$ out of all the edges between $j^{-}$
and its neighboring out-leaves are randomly chosen and matched, then $z_{j^{-}}$ is set to $0$;
correspondingly,
$j^{-}$ is removed along with all its adjacent edges.
(3) For an in-leaf, say $j^{-}$,
all the edges between $j^{-}$ and its neighboring out-roots are matched, then $z_{j^{-}}$ is set to $0$;
correspondingly,
$j^{-}$ and all its neighboring out-roots
are removed along with all their adjacent edges.
The above steps are locally optimal for constructing a $z$-matching solution,
and a simple explanation is left in appendix A.
We also show that, even there is some randomness in the order of removing leaves and roots defined as above, the vertices in the final core structure are independent of sequences of removal orders. We leave the proof in appendix B.

On a graph instance, if GLRB leaves no core, the matched edges constitute an approximate $z$-matching solution.
If GLRB leaves a core, a heuristic or statistical mechanics-based method can be further applied on it until a $z$-matching configuration is found.
In this paper, we adopt a simple heuristic method: an out-vertex in a core with the smallest degree is chosen, say $i^{+}$, and an edge adjacent to it, say $(i^{+}, j^{-})$ with $j^{-} \in \partial i^{+}$,
is randomly chosen and matched, then $z_{j^{-}}$ is reduced by $1$; correspondingly, $i^{+}$ is removed and $j^{-}$ is also removed if $z_{j^{-}} = 0$.
GLRB and the heuristic steps on core are carried out alternately on a reducing subgraph, until all edges are removed. The set of all matched edges corresponds to an approximate $z$-matching solution.
See figures \ref{fig:model} (d)-(i) for an illustration.

Here we estimate the time complexity of GLRB and the heuristic method for $z$-matchings.
In our realization of data structures of graph topology and dynamical processes on a graph, we keep the removal of each vertex based on its degree and its influence on nearest neighbors in a local sense.
The initial traversal of a graph assigns degrees for each vertex and keeps a record of sets of leaves and roots, which takes a time complexity of $|V_{+}| + |V_{-}| + |E|$.
In each step of removal in GLRB, a leaf or a root is marked as being removed, degrees of its neighbors are reduced by one, and the sets of leaves and roots are updated correspondingly. The time complexity of this single step is generally the degree of a vertex to be removed plus $1$ for the vertex per se. Correspondingly, the iterative procedure in GLRB takes a time complexity of the size of edges and vertices not in the final core structure, whose largest size is simply $|V_{+}| + |V_{-}| + |E|$.
Summing the above two terms, GLRB on a sparse graph takes a linear time complexity of its vertex size.
In the heuristic method to construct a $z$-matching, a greedy step on a core structure involves selecting and removing an edge. The extra overhead to remove an edge in a core mainly derives from finding out-vertices with smallest degrees, which can be considered as sublinear of out-vertex size or even constant if we keep a dynamic record of the smallest degree for out-vertices and the corresponding out-vertex set.
The algorithm can further speed up if multiple out-vertices are removed in a single heuristic step.
Considering that all vertices and edges are finally removed either in GLRB or in greedy steps to find a $z$-matching, the heuristic method for $z$-matching on a sparse graph still has a linear time complexity of vertex size, yet with a larger multiplicative factor than that for GLRB.

\section{Theory}
\label{sec:theory}

\subsection{Analytical framework for percolation and $z$-matching sizes}

Our local algorithm to approximate a $z$-matching configuration applies on any bipartite graph instance. Yet on uncorrelated random bipartite graphs, we can develop a mean-field theory to analytically describe GLRB and further estimate $z$-matching sizes.
By uncorrelated random bipartite graphs, we mean random bipartite graphs with no correlation among in-degrees, nor among out-degrees, nor between in-degrees and out-degrees.

Before establishing our analytical framework,
we present notations of degree distributions for uncorrelated graph ensembles.
On a bipartite graph $B = \{V_{+}, V_{-}, E\}$,
the out-degree distribution $P_{+}(k_{+})$
is the probability of randomly finding an out-vertex in $V_{+}$ with a degree $k_{+}$.
The mean degree of out-vertices in $B$ is $c_{+} = \sum _{k_{+}} k_{+} P_{+}(k_{+})$.
The in-degree distribution $P_{-} (k_{-})$ and the mean degree of in-vertices $c_{-}$ for $V_{-}$
can be defined in a similar way.
It is easy to see that $|V_{+}| c_{+} = |V_{-}| c_{-} = |E|$.
On a randomly chosen edge $(i^{+}, j^{-}) \in E$
between an out-vertex $i^{+}$ and an in-vertex $j^{-}$,
following $j^{-}$ to $i^{+}$,
we define the excess out-degree distribution $Q_{+}(k_{+})$
as the probability that $i^{+}$ has a degree $k_{+}$;
following $i^{+}$ to $j^{-}$,
we define the excess in-degree distribution $Q_{-}(k_{-})$
as the probability that $j^{-}$ has a degree $k_{-}$.
We can see that $Q_{\pm} (k_{\pm}) = k_{\pm} P_{\pm} (k_{\pm}) / c_{\pm}$.

We follow the language of cavity method
\cite{Mezard.Montanari-2009}
in establishing our mean-field theory of GLRB.
The general procedure of a cavity method for a percolation problem on graphs
is to define cavity probabilities describing state evolution in the percolation process,
derive their coupled equations,
calculate their stable fixed solutions,
and finally compute properties of graphical substructure revealed by the percolation process.
Pertinent properties of GLRB that we are interested in are
the relative sizes of out-(in-)vertices in a core $n^{+}$($n^{-}$) normalized by $|V^{+}|$($|V^{-}|$),
edges in a core $l$ normalized by $|V^{+}|$,
and matched edges $w$ normalized by $|V^{+}|$.
For GLRB on a random bipartite graph $B$,
an edge $(i^{+}, j^{-}) \in E$ between an out-vertex $i^{+}$ and an in-vertex $j^{-}$ is randomly chosen,
and on it we define a set of four cavity probabilities $\{\alpha ^{+}, \beta ^{-}, \alpha ^{-}, \beta ^{+}\}$.
From $j^{-}$ to $i^{+}$, we define $\alpha ^{+}$ as the probability of $i^{+}$ to be an out-leaf, with $j^{-}$ as its corresponding in-root.
From $i^{+}$ to $j^{-}$, we define $\beta ^{-}$ as the probability of $j^{-}$ to be an in-root, with $i^{+}$ not among its corresponding out-leaves.
Also from $i^{+}$ to $j^{-}$, we define $\alpha ^{-}$ as the probability of $j^{-}$ to be an in-leaf, with $i^{+}$ as one of its corresponding out-roots.
From $j^{-}$ to $i^{+}$, we define $\beta ^{+}$ as the probability of $i^{+}$ to be an out-root, with $j^{-}$ not among its corresponding in-leaves.
Thus, from $j^{-}$ to $i^{+}$, the probability of $i^{+}$ to be in a core is $1 - \alpha ^{+} - \beta ^{+}$. Likewise, from $i^{+}$ to $j^{-}$, the probability of $j^{-}$ to be in a core is $1 - \alpha ^{-} - \beta ^{-}$.

It is easy to see that, whether a vertex becomes one of the four local structures depends on the local states of its nearest neighbors in removal process.
Besides, on large sparse random graphs, we adopt a locally tree-like structure assumption for mean-field theory \cite{Mezard.Montanari-2009}, which states that on a graph $B \backslash i^{+}$ (or $B \backslash j^{-}$) with $i^{+}(j^{-})$ removed along with all its adjacent edges,
the states of vertices in $\partial i^{+}$ ($\partial j^{-}$)
are independent of each other in a dynamical process on graphs due to long paths between them.
Based on these two ingredients, we establish self-consistent equations of four cavity probabilities as
\begin{eqnarray}
\label{eq:alpha_p}
\alpha ^{+}
&&
= \sum _{k_{+} = 1}^{+ \infty}
Q_{+}(k_{+}) (\beta ^{-})^{k_{+} - 1}, \\
\label{eq:beta_m}
\beta ^{-}
&&
= \sum _{k_{-} = z}^{+ \infty}
Q_{-} (k_{-}) \sum _{s = z}^{k_{-} - 1}
{k_{-} - 1 \choose s}
(\alpha ^{+})^{s}
(1 - \alpha ^{+})^{k_{-} - 1 - s}, \\
\label{eq:alpha_m}
\alpha ^{-}
&&
= \sum _{k_{-} = 1}^{+ \infty}
Q_{-} (k_{-}) \sum _{s = 0}^{z - 1}
{k_{-} - 1 \choose s}
(1 - \beta ^{+})^{s}
(\beta ^{+})^{k_{-} - 1 - s}, \\
\label{eq:beta_p}
\beta ^{+}
&&
= \sum _{k_{+} = 1}^{+ \infty}
Q_{+}(k_{+}) [1 - (1 - \alpha ^{-}) ^{k_{+} - 1}].
\end{eqnarray}
With the physical stable solutions of $\{\alpha ^{+}, \beta ^{-}, \alpha ^{-}, \beta ^{+}\}$,
we can calculate $\{n^{+}, n^{-}, l, w\}$ as
\begin{eqnarray}
\label{eq:n_p}
n^{+}
&&
= \sum _{k_{+} = 2}^{+ \infty} P_{+}(k_{+})
\sum _{s = 2}^{k_{+}}
{k_{+} \choose s}
(1 - \alpha ^{-} - \beta ^{-})^{s}
(\beta ^{-})^{k_{+} - s}, \\
\label{eq:n_m}
n^{-}
&&
= \sum _{k_{-} = z + 1}^{+ \infty} P_{-}(k_{-})
\sum _{t = 0}^{z - 1}
{k_{-} \choose t}
(\alpha ^{+})^{t}
\sum _{s = z + 1 - t}^{k_{-} - t}
{k_{-} - t \choose s}
(1 - \alpha ^{+} - \beta ^{+})^{s}
(\beta ^{+})^{k_{-} - t - s}, \\
\label{eq:l}
l && = c_{+} (1 - \alpha ^{+} - \beta^{+}) (1 - \alpha ^{-} - \beta^{-}), \\
\label{eq:w}
w
&&
= \sum _{k_{+} = 0}^{+ \infty} P_{+}(k_{+})
[1 - (1 - \alpha ^{-})^{k_{+}}]
+ \phi z \sum _{k_{-} = z}^{+ \infty}
P_{-}(k_{-}) \sum _{s = z}^{k_{-}}
{k_{-} \choose s}
(\alpha ^{+})^{s}
(1 - \alpha ^{+})^{k_{-} - s} \nonumber \\
&&
+ \phi \sum _{k_{-} = z + 1}^{+ \infty} P_{-}(k_{-})
\sum _{t = 1}^{z - 1} t
{k_{-} \choose t}
(\alpha ^{+})^{t}
\sum _{s = z + 1 - t}^{k_{-} - t}
{k_{-} - t \choose s}
(1 - \alpha ^{+} - \beta ^{+})^{s}
(\beta ^{+})^{k_{-} - t - s} \nonumber \\
&&
- \phi z \sum _{k_{-} = z}^{+ \infty} P_{-}(k_{-})
{k_{-} \choose z}
(\alpha ^{+})^{z}
(\beta ^{+})^{k_{-} - z}.
\end{eqnarray}

We first briefly explain the self-consistent equations by considering four probabilities on an edge $(i^{+},j^{-}) \in E$ as in their definition.
For equation (\ref{eq:alpha_p}), if $i^{+}$ becomes an out-leaf,
from $i^{+}$ to its nearest neighbors $\partial i^{+} \backslash j^{-}$, all these in-vertices are removed as in-roots, while $i^{+}$ is not any of their corresponding out-leaves.
For equation (\ref{eq:beta_m}), if $j^{-}$ becomes an in-root,
from $j^{-}$ to its nearest neighbors $\partial j^{-} \backslash i^{+}$, there are at least $z$ out-leaves, since $i^{+}$ is not any of $j^{-}$'s out-leaves.
For equation (\ref{eq:alpha_m}), if $j^{-}$ becomes an in-leaf,
from $j^{-}$ to its nearest neighbors $\partial j^{-} \backslash i^{+}$, there are at most $z - 1$ out-vertices left after removal of out-roots, since $i^{+}$ is already among $j^{-}$'s out-roots.
For equation (\ref{eq:beta_p}), if $i^{+}$ becomes an out-root,
from $i^{+}$ to its nearest neighbors $\partial i^{+} \backslash j^{-}$, there are at least one in-leaf for $i^{+}$, since $j^{-}$ is not an in-leaf of $i^{+}$.
To write down the four self-consistent equations for probabilities in current forms, we follow a typical formulation in percolation theories on random graphs: we first calculate a single probability term for proper states of the nearest neighbors for a vertex with a given degree $k$, and then sum terms under all $k$ with excess degree distribution $Q(k)$ or degree distribution $P(k)$.
These products of combinatorial and power terms aim to enumerate all the possibilities of feasible local configurations of nearest neighbors of a vertex under a certain constraint.
For example, on the right-hand side of equation (\ref{eq:alpha_p}), for a given $k_{+}$ of $i^{+}$, all its nearest neighbors $\partial i^{+} \backslash j^{-}$ as in-roots has a probability of $(\beta ^{-})^{k_{+} - 1}$, and then we sum these probability terms under all $k^{+}$ with an excess out-degree distribution $Q_{+}(k_{+})$.
On the right-hand side of equation (\ref{eq:beta_m}), the nearest neighbors $\partial j^{-} \backslash i^{+}$ of $j^{-}$ can be out-leaves or not. For a given $k_{-}$ of $j^{-}$, we simply calculate the probability of proper configurations with $s$ out-leaves and $k_{-} - 1 - s$ non-out-leaves with a combinatorial coefficient under a local constraint $s \geqslant z$, and then sum such terms under all $k_{-}$ with an excess in-degree distribution $Q_{-}(k_{-})$. In the following equations, there are more complex situations, in which three types of states are possible for a proper configuration of nearest neighbors of a vertex. Yet the logic of derivation for those equations remains the same.

We then explain the three equations for core structure.
For equation (\ref{eq:n_p}), we consider a randomly chosen out-vertex $i^{+} \in V_{+}$.
If $i^{+}$ is in a core, from $i^{+}$ to its nearest neighbors $\partial i^{+}$, there are three conditions to be satisfied. First, there is no in-leaf, or $i^{+}$ will become an out-root. Then, there can be in-roots. Finally, there are at least two vertices which are also in the core, or $i^{+}$ will become an out-leaf or trivially isolated.
For equation (\ref{eq:n_m}), we consider a randomly chosen in-vertex $j^{-} \in V_{-}$.
If $j^{-}$ is in a core, from $j^{-}$ to its nearest neighbors $\partial j^{-}$, there are also three conditions to be satisfied. First, there are at most $z - 1$ out-leaves, or $j^{-}$ will become an in-root. Then, there can be out-roots. Finally, there are nearest neighbors also in the core, whose size along with that of $j^{-}$'s out-leaves is $\geqslant z+ 1$, or $j^{-}$ will become an in-leaf.
For equation (\ref{eq:l}), we consider a randomly chosen edge $(i^{+}, j^{-}) \in E$.
If $(i^{+}, j^{-})$ is in a core, from $i^{+}$ to $j^{-}$, $j^{-}$ must be in the core. At the same time, from $j^{-}$ to $i^{+}$, $i^{+}$ must be also in the core. The probability is then multiplied with the edge density normalized by out-vertex size $|V_{+}|$, which is simply $c_{+}$.

Finally, we consider equation (\ref{eq:w}) for the relative size of matched edges from GLRB.
The first term counts the contribution of matched edges from out-roots, in which each out-root contributes to a $z$-matching a matched edge between the out-root and one of its in-leaves.
We consider a randomly chosen out-vertex $i^{+} \in V_{+}$. If $i^{+}$ is an out-root, from $i^{+}$ to its nearest neighbors $\partial i^{+}$, there are at least one in-leaf.
The second term counts the contribution from in-roots, in which each in-root contributes exactly $z$ matched edges connecting the in-root and its out-leaves.
We consider a randomly chosen in-vertex $j^{-} \in V_{-}$.
If $j^{-}$ is an in-root, from $j^{-}$ to its nearest neighbors $\partial j^{-}$, there are at least $z$ out-leaves.
The third term counts the contribution from in-vertices in the core, whose derivation is quite similar with equation (\ref{eq:n_m}).
Consider a vertex $j^{-} \in V_{-}$, 
if $j^{-}$ is in the core, from $j^{-}$ to its nearest neighbors $\partial j^{-}$, there are $t \in [0, z - 1]$ out-leaves whose adjacent edges will be matched. For a given $k_{-}$, the relative size of matched edges contributed by $j^{-}$ can be calculated by multiplying $t$ with the probability of such a local configuration of $j^{-}$. Then such terms are summed with all $k_{-}$ with probability $P_{-}(k_{-})$.
The fourth term deals with a double-counted case when an in-vertex is both an in-root and an in-leaf,
while an in-vertex of such type contributes $z$ matched edges to a $z$-matching.
Consider a vertex $j^{-} \in V_{-}$, if $j^{-}$ is such an in-vertex, from $j^{-}$ to its nearest neighbors $\partial j^{-}$, there are exactly $z$ out-leaves, while the other neighbors are simply out-roots which will be removed.
In the last three terms, the coefficient $\phi$  is a readjusting factor, since terms of $w$ are normalized with the out-vertex size $|V_{+}|$ rather than the in-vertex size $|V_{-}|$.

We denote $y$ as the fraction of a $z$-matching or the relative size of edges in it normalized by $|V^{+}|$.
If GLRB leaves no core, we simply have $y = w$.
If GLRB leaves a core, $y$ consists of $w$ from iterative removal in GLRB and a term on the matched edges from the core. Estimating the exact size of matched edges in a core is beyond the capability of GLRB and its theoretical framework here, yet we can still reach an approximation with the current probabilistic framework of percolation under an assumption of perfect matching of cores.
Under an extension of the perfect matching of cores on a single undirected graph
\cite{Karp.Sipser-IEEFoCS-1981},
we assume that in a core after GLRB on a large bipartite graph, with proper algorithms either all the out-vertices or all the in-vertices can be fully matched. By fully matched out-(in-)vertices we mean that an out-(in-)vertex being adjacent to $1$($z$) matched edge(s).
We can see that the size of matched edges in a core can be simply estimated as the smaller one of the maximal size of matched edges adjacent to out-vertices in a core, which is denoted by $n^{+}$, and the maximal size of matched edges that the in-vertices in a core can provide.
We define $n_{\rm M}^{-}$ as the relative size of maximal edges normalized by $|V^{-}|$ the in-vertices in a core can match.
It is easy to see that, each in-vertex in a core has $t \in [0, z - 1]$ neighboring out-leaves, thus it can further match at most $z - t$ edges also in the core. $n_{\rm M}^{-}$ has a quite similar logic with deriving equation (\ref{eq:n_m}) and the third term of equation (\ref{eq:w}). We have
\begin{eqnarray}
\label{eq:n_my}
n_{\rm M}^{-}
&&
= \sum _{k_{-} = z + 1}^{+ \infty} P_{-}(k_{-})
\sum _{t = 0}^{z - 1} (z - t)
{k_{-} \choose t}
(\alpha ^{+})^{t}
\sum _{s = z + 1 - t}^{k_{-} - t}
{k_{-} - t \choose s}
(1 - \alpha ^{+} - \beta ^{+})^{s}
(\beta ^{+})^{k_{-} - t - s}.
\end{eqnarray}
Finally, based on the percolation analysis of GLRB and the assumption of perfect matchings on a core after GLRB, we estimate $y$ as
\begin{eqnarray}
\label{eq:y}
y = w + \min \{n^{+}, \phi n_{\rm M}^{-}\}.
\end{eqnarray}
Equations (\ref{eq:alpha_p}) - (\ref{eq:y})
constitute our analytical framework of GLRB and $z$-matchings on random bipartite graphs.
We modify the above equations into a form explicit for numerical computation, and their final form is left in appendix C.
With $z = 1$ and $\phi = 1$, these equations reduce to the theory for the MM problem on directed graphs
\cite{Zhao.Zhou-PRE-2019}.

\subsection{Comparison with Krea\v{c}i\'{c} and Bianconi's theory}

Before our analytical study of $z$-matchings here,
authors in \cite{Kreacic.Bianconi-EPL-2019}
present a belief propagation algorithm,
whose zero-temperature limit also gives an analytical prediction of fractions of $z$-matchings
on random bipartite graphs.
Here we discuss differences between these two frameworks.
After substituting
$N$, $M$, $\langle k \rangle$, $\langle q \rangle$, $C/N$, $w_{1}$, $w_{2}$, $\vec{w}_{1}$, $\vec{w}_{2}$
in \cite{Kreacic.Bianconi-EPL-2019} with
$|V_{+}|$, $|V_{-}|$, $c_{+}$, $c_{-}$, $y$, $\alpha ^{+}$, $\beta ^{+}$, $\alpha ^{-}$, $\beta ^{-}$
in our framework respectively,
we can see that equations (24) and (25) in \cite{Kreacic.Bianconi-EPL-2019}
are simply equations (\ref{eq:alpha_p}) - (\ref{eq:beta_p}) here.
Also in our notations, after some modification, their prediction of $z$-matching fraction $y_{\rm KB}$ is
\begin{eqnarray}
\label{eq:y_KB}
2 y_{\rm KB}
\label{eq:y_KB}
&& = \phi z + 1
- \sum _{k_{+} = 0}^{+\infty} P_{+} (k_{+}) (\beta ^{-})^{k_{+}}
+ \sum _{k_{+} = 0}^{+\infty} P_{+} (k_{+}) [1 - (1 - \alpha ^{-})^{k_{+}}] \nonumber \\
&&
+ \phi z \sum _{k_{-} = 0}^{+\infty} P_{-} (k_{-})
\sum _{s = z}^{k_{-}}
{k_{-} \choose s}
(\alpha ^{+})^{s} (1 - \alpha ^{+})^{k_{-} - s} \nonumber \\
&&
+ \phi \sum _{k_{-} = 0}^{+\infty} P_{-} (k_{-})
\sum _{t = 1}^{z - 1}
t {k_{-} \choose t}
(\alpha ^{+})^{t} (1 - \alpha ^{+})^{k_{-} - t} \nonumber \\
&&
- \phi \sum _{k_{-} = 0}^{+\infty} P_{-} (k_{-})
\sum _{t = 0}^{z - 1}
{k_{-} \choose t} (\alpha ^{+})^{t} 
\sum _{s = 0}^{z - t - 1}
{k_{-} - t \choose s}
(z - t - s) (1 - \alpha ^{+} - \beta ^{+})^{s} (\beta ^{+})^{k_{-} - t - s} \nonumber \\
&&
- c_{+} [\alpha ^{+} (1 - \beta^{-}) + \alpha ^{-} (1 - \beta ^{+})].
\end{eqnarray}
%

As it is not explicit to find a consistent geometric interpretation for each term of equation (\ref{eq:y_KB}), we compare it with our prediction of $y$ only in specific cases.
As a simple case, we consider $\phi = 1$, $z = 1$, and $P_{+}(k_{+}) = P_{-}(k_{-})$.
This case corresponds to the MM problem on directed graphs with the same out- and in-degree distributions
\cite{Liu.Csoka.Zhou.Posfai-PRL-2012},
in which there is only continuous core percolation.
In our framework, we have
$\alpha ^{+} = \alpha ^{-}$, $\beta ^{+} = \beta ^{-}$, $n^{+} = n^{-}$, $n^{-}_{\rm M} = n^{-}$.
We conveniently define
$P_{\pm} (k_{\pm}) \equiv P(k)$,
$c_{\pm} = c$,
$\alpha ^{\pm} \equiv \alpha$,
$\beta ^{\pm} \equiv \beta$, and
$n^{\pm} \equiv n$.
Based on equation (\ref{eq:y}) in our framework, we have $y = w + n$.
After some derivation,
we see that two frameworks both predict $y$ as
\begin{eqnarray}
y = \sum _{k = 0}^{+\infty} P(k) [2 - (1 - \alpha)^{k} - \beta ^{k}] - c \alpha (1 - \beta).
\end{eqnarray}
%

Generally on the methodological side,
the Max-Sum equations of the cavity method at zero-temperature limit in
\cite{Kreacic.Bianconi-EPL-2019}
are equivalent to the iterative equations in the percolation analysis of GLRB in our framework.
The reason is that at zero-temperature limit, to minimize local energy terms,
vertices can be assigned with a certain state (being matched or not)
through incoming biased cavity fields ($\{-1, +1\}$)
based on local information of graphical structure,
which is logically similar to the local optimal steps in our framework.
In the regime where there is no percolation,
these two frameworks, if derived properly, lead to similar predictions.
For the $z$-matching problem here, when there is no percolation,
we have $1 - \alpha ^{\pm} - \beta ^{\pm} = 0$.
After some derivation,
both equation (\ref{eq:y_KB}) from \cite{Kreacic.Bianconi-EPL-2019}
and equation (\ref{eq:w}) from our framework reduces to the same formula for $y$ as
\begin{eqnarray}
&& y
= \sum _{k_{+} = 0}^{+ \infty} P_{+}(k_{+})
[1 - (1 - \alpha ^{-})^{k_{+}}]
+ \phi z \sum _{k_{-} = z + 1}^{+ \infty}
P_{-}(k_{-}) \sum _{s = z + 1}^{k_{-}}
{k_{-} \choose s}
(\alpha ^{+})^{s}
(1 - \alpha ^{+})^{k_{-} - s}.
\end{eqnarray}
It is easy to see that, the right-hand side of the above formula consists of two terms,
one as the fraction of out-roots
and the other one as the fraction of in-roots, each of which contributes $z$ matched edges
among all edges adjacent to its $\geqslant z + 1$ neighboring out-leaves.

Yet in the regime where there is a percolation,
especially a discontinuous one or with unequal core sizes ($n_{+} \neq n_{-}$),
the calculation in belief propagation formulation at zero-temperature limit can be tricky,
as both the cavity and marginal probabilities of a vertex being matched or not
is described by trivial cavity fields $0$ in an average sense.
In the Results section, we show predictions from two frameworks on a specific random bipartite graph ensemble.
We will see that, in the regime with discontinuous percolation transitions,
the theory in \cite{Kreacic.Bianconi-EPL-2019} mostly makes unphysical predictions.
Thus based on a geometric and probabilistic interpretation,
our framework instead can offer a more reasonable and consistent prediction.

\subsection{Percolation analysis of $k$-XORSAT problem}

As our local method and analytical framework are defined on bipartite graphs,
it has potential in approximating other optimization problems defined on factor graphs or hypergraphs.
A typical example is the $k$-XORSAT problem\cite{RicciTersenghi.Weigt.Zecchina-PRE-2001}.
A random $k$-XORSAT problem involves finding an assignment to $N$ binary variables (\{0, 1\}) to satisfy given $M = \alpha N$ ($\alpha$ as constraint density) parity checks, each of which is based on $k$ variables randomly chosen from the $N$ variables. It can be easily mapped to the $p$-spin model studied in statistical physical literature. In a factor graph representation, parity checks and variables can be denoted as factor and variable nodes respectively, and their inclusion relation can be denoted as an edge connecting a factor node and corresponding variable nodes.

For $k$-XORSAT problem,
there are statistical physical approaches in replica symmetric and first-step replica symmetry breaking formalisms
\cite{RicciTersenghi.Weigt.Zecchina-PRE-2001,Franz.etal-PRL-2001}.
Also, a leaf removal algorithm on hypergraphs
\cite{
Cocco.etal-PRL-2003,
Mezard.RicciTersenghi.Zecchina-JStatPhys-2003}
is adopted as a rigorous graphical method
to locate transition thresholds and describe the entropy landscapes of solutions.
In the original analysis of leaf removal process on hypergraphs,
the underlying core percolation problem is solved based on dynamical evolutions of degree distributions.

In the core percolation analysis of $k$-XORSAT problem in
\cite{
Cocco.etal-PRL-2003,
Mezard.RicciTersenghi.Zecchina-JStatPhys-2003},
a factor node has a uniform degree $k$ and the variable nodes follow a Poisson distribution.
Here, we show that our GLRB provides a percolation analysis of XORSAT problems with arbitrary uncorrelated degree distributions of factor and variable nodes.
In the language of our framework, the leaf removal process in \cite{
Cocco.etal-PRL-2003,
Mezard.RicciTersenghi.Zecchina-JStatPhys-2003}
assigns a state to an out-vertex with only one neighboring in-vertex and further removes this neighbor along with all its adjacent edges. This algorithm simply corresponds to removal procedures in GLRB on bipartite graphs with only out-leaves and in-roots and also $z = 1$.
Based on  equations (\ref{eq:alpha_p}) - (\ref{eq:w}) in our theoretical framework for GLRB,
by setting $\alpha ^{-} = \beta ^{+} = 0$,
we have a theoretical framework for the core percolation induced by the leaf removal algorithm for XORSAT problems on uncorrelated hypergraphs with degree distribution $P_{+}(k_{+})$ for variable nodes and degree distribution $P_{-}(k_{-})$ for factor nodes as
\begin{eqnarray}
\alpha ^{+}
\label{eq:xorsat_a}
&& = \sum _{k_{+} = 1}^{+ \infty} Q_{+} (k_{+}) (\beta ^{-})^{k_{+} - 1}, \\
\beta^{-}
\label{eq:xorsat_b}
&& = \sum _{k_{-} = 1}^{+ \infty} Q_{-}(k_{-}) [1 - (1 - \alpha ^{+})^{k_{-} - 1}],\\
n^{+}
&&= \sum _{k_{+} = 2}^{+ \infty} P_{+} (k_{+})
\sum _{s = 2}^{k_{+}} {k_{+} \choose s} (1 - \beta^{-})^{s} (\beta ^{-})^{k_{+} - s},\\
n^{-}
&&= \sum _{k_{-} = 2}^{+ \infty} P_{-}(k_{-}) (1 - \alpha ^{+})^{k_{-}},\\
l
&& = c_{+} (1- \alpha^{+}) (1 - \beta^{-}),\\
w
\label{eq:xorsat_w}
&&= \phi \sum _{k_{-} = 1}^{+ \infty} P_{-}(k_{-}) [1 - (1 - \alpha ^{+})^{k_{-}}] - \phi P_{-}(1) \alpha ^{+}.
\end{eqnarray}
We should mention that equation (\ref{eq:xorsat_a}) is simply equation (\ref{eq:alpha_p}), and here $w$ is the relative size of variable nodes assigned with certain states in the leaf removal process.
For a typical $k$-XORSAT problem, we simply consider a uniform in-vertex degree as $P_{-}(k_{-}) = \delta (k_{-}, k)$.

Following the analysis on thresholds in 
\cite{
Cocco.etal-PRL-2003,
Mezard.RicciTersenghi.Zecchina-JStatPhys-2003},
we know that for $k \geqslant 3$ the discontinuous emergence of $n_{+}$ and $n_{-}$ signifies clustering transitions at $\phi _{d}$, and the critical constraint density $\phi _{s}$ satisfying $\phi _{s} n_{-}(\phi _{s}) = n_{+}(\phi _{s})$ sets SAT/UNSAT thresholds.
With our framework here, for a large random factor graph with degree distributions $P_{+}(k_{+})$ and $P_{-}(k_{-})$ for variable and factor nodes respectively, we can easily derive these two thresholds and also energy and entropy profiles for ground state configurations.

In short, our framework generalizes the previous percolation analysis of a leaf removal algorithm onto factor graphs with factor and variable nodes described by arbitrary uncorrelated degree distributions, thus pushes the power of this algorithm into a more general context of factor graphs.

\section{Results}
\label{sec:results}

\begin{figure*}
\begin{center}
 \includegraphics[width = 0.99 \linewidth]{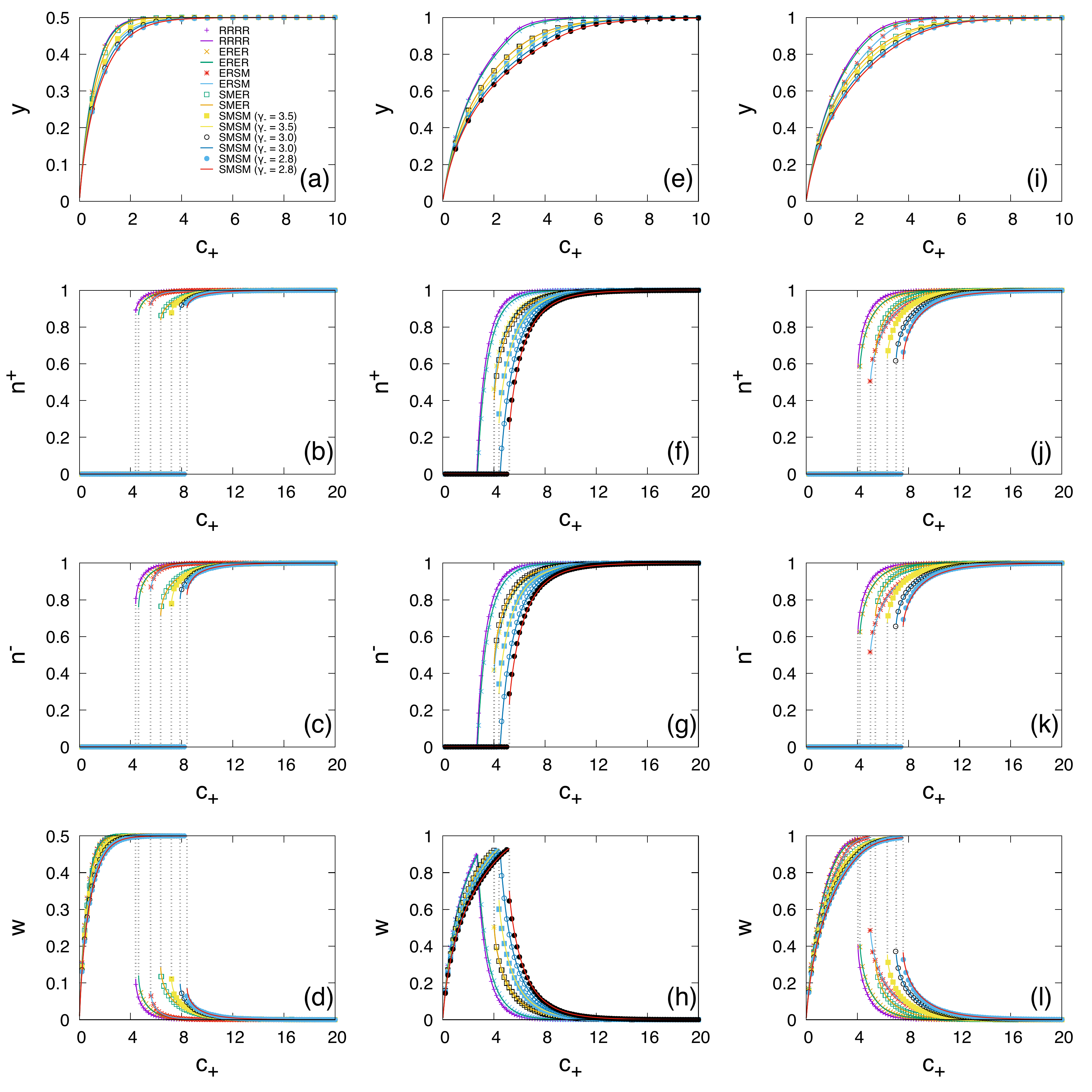}
\end{center}
\caption{
\label{fig:rand_z1}
$z$-matchings and GLRB on random bipartite graphs with $z = 1$.
Details of graph ensembles are in the main text.
In (a)-(d), (e)-(h), and (i)-(l),
results in the cases of $\phi = 0.5, 1.0, 1.2$ are shown respectively.
In the local algorithm on cores,
we remove a fraction $f$ or a minimal size $N_{\rm min}$ of out-vertices to speed up the algorithm.
Here we set $f = 0.001$ and $N_{\rm min} = 1$.
Each sign is a simulation result on a graph instance with an out-vertex size $|V_{+}| = 10^5$.
Solid lines are for analytical results on infinitely large graphs.
Dashed vertical lines denote discontinuous changes in analytical results.}
\end{figure*}
\begin{figure*}
\begin{center}
 \includegraphics[width = 0.99 \linewidth]{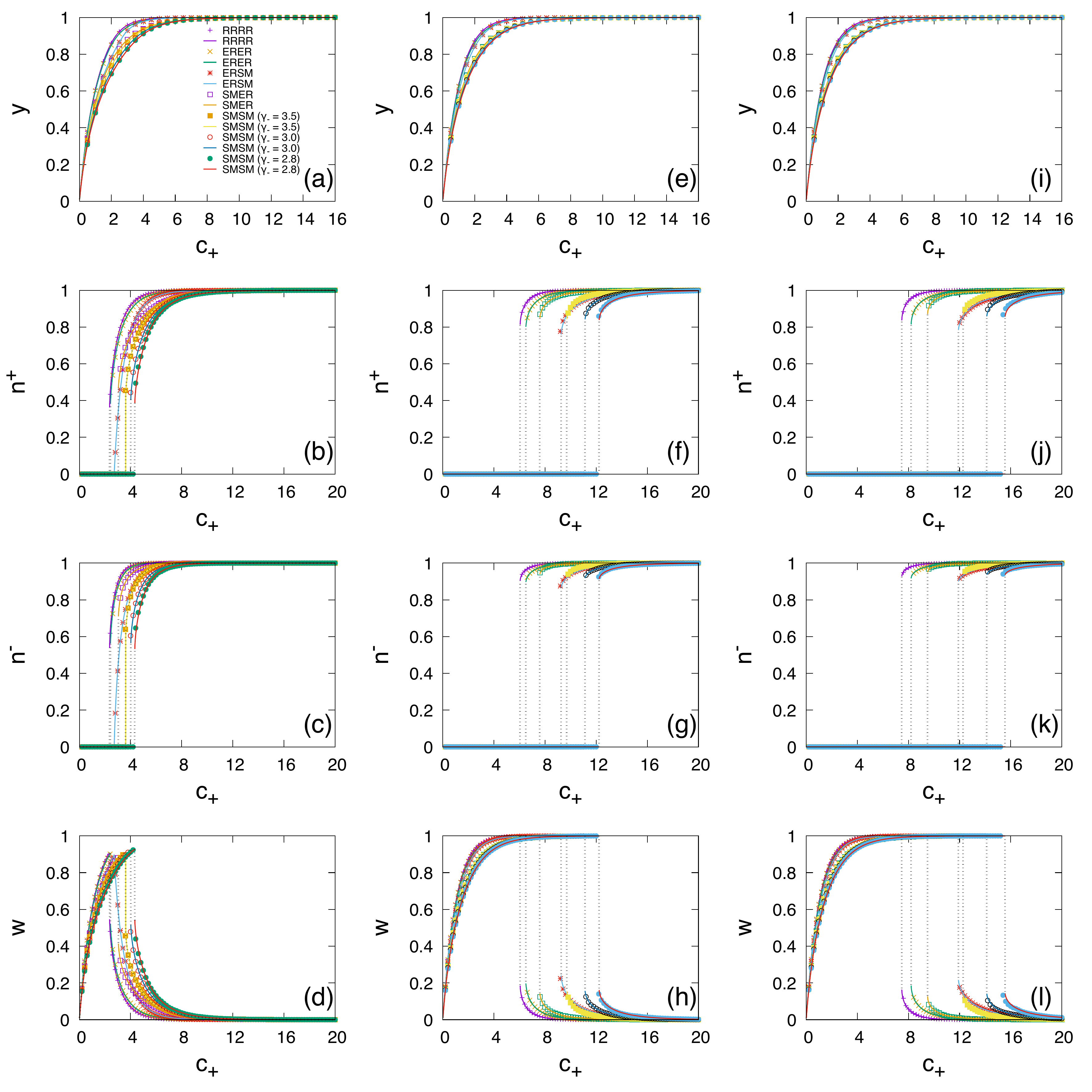}
\end{center}
\caption{
\label{fig:rand_z2}
Same with figure \ref{fig:rand_z1} yet with $z = 2$.}
\end{figure*}
\begin{figure*}
\begin{center}
 \includegraphics[width = 0.99 \linewidth]{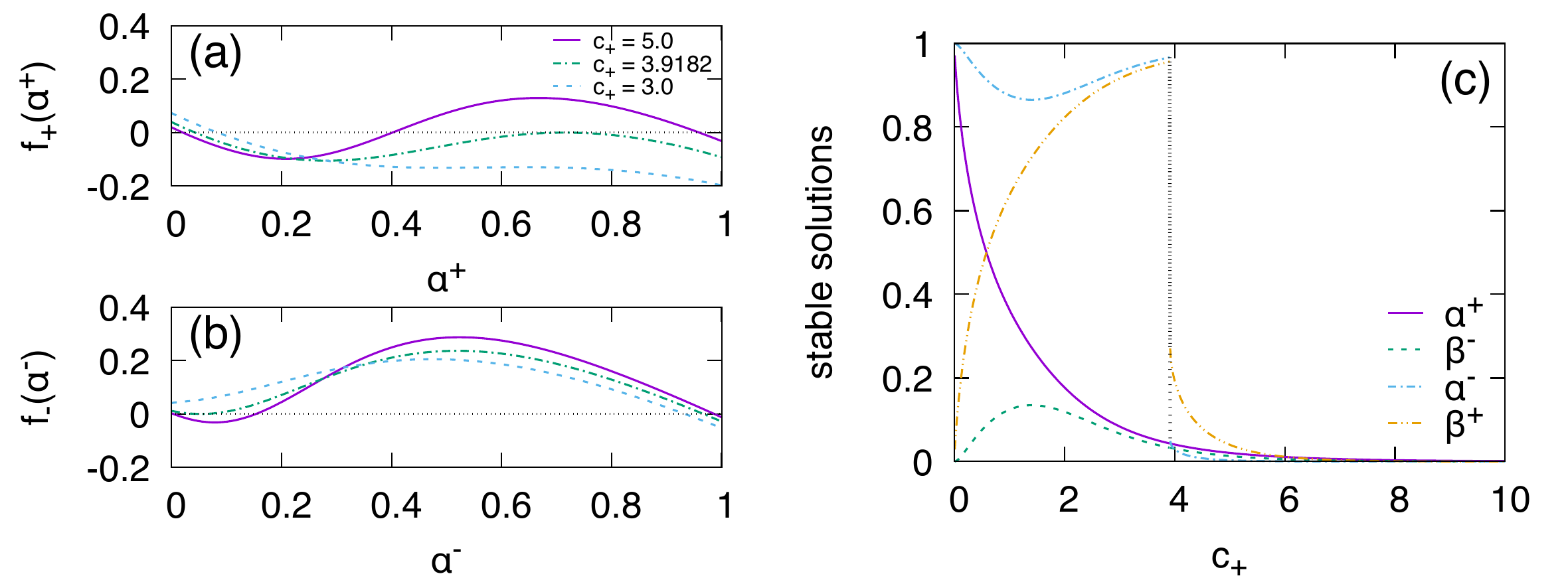}
\end{center}
\caption{
\label{fig:fixed}
Analysis of fixed solutions of $(\alpha ^{+}, \beta^{-}, \alpha ^{-}, \beta ^{+})$
in the percolation theory for GLRB.
The underlying graph ensembles are infinitely large SMER graphs with $\gamma _{+} = 3.0$.
We set $\phi = 0.6$ and $z = 2$.
(a), (b) The behavior of $f_{+} (\alpha ^{+})$ and $f_{-} (\alpha ^{-})$, respectively.
(c) The evolution of stable fixed solutions of $(\alpha ^{+}, \beta^{-}, \alpha ^{-}, \beta ^{+})$
with increasing $c_{+}$.
Dotted vertical lines denote discontinuous changes.}
\end{figure*}

\subsection{Result on random graphs}

Here, we test our local algorithm and analytical theory for the $z$-matching problem
on some random bipartite graph models.

We consider some typical degree profiles for out- and in-vertices of random bipartite graphs.
First we consider a diluted version of uniform degree distributions shown in random regular (RR) graphs.
On a large bipartite graph with a uniform out-(in-)degree $k_{0} \geqslant 2$,
to construct a diluted graph instance with heterogeneous degrees,
a fraction $1 - \rho$ ($\rho \in [0, 1]$ as the dilution parameter) of its edges are randomly chosen and removed.
The mean out-(in-) degree is $c = \rho k_{0}$.
Thus we have a diluted bipartite graph with an out-(in-)degree distribution as
\begin{eqnarray}
P(k) = {k_{0} \choose k} \rho ^k (1 - \rho) ^{k_{0} - k}.
\end{eqnarray}
%
Then we consider Poisson degree distributions in ER random graphs
\cite{
Erdos.Renyi-PublMath-1959,
Erdos.Renyi-Hungary-1960}.
The Poisson distribution for out-(in-)vertex degrees is
\begin{eqnarray}
P(k) = {\rm e}^{- c} \frac {c^k}{k!},
\end{eqnarray}
where $c$ is the mean out-(in-)degree.
We further consider power-law distributions
\cite{Barabasi.Albert-Science-1999}
for out-(in-)vertex degrees as $P(k) \propto k^{- \gamma}$ with $\gamma$ as the out-(in-)degree exponent in scale-free networks. Here we adopt the static model
\cite{
Goh.Kahng.Kim-PRL-2001,
Catanzaro.PastorSatorras-EPJB-2009,
Lee.Goh.Kahng.Kim-EPJB-2006}
to construct asymptotical power-law degree profiles for graphs.
With only $\gamma$, an out-(in-)vertex size $N$, and a mean out-(in-)degree $c$ as parameters, the static model can generate a network instance with a broad degree range encompassing naturally small degrees, which makes GLRB possible without diluting a graph instance.
The out-(in-)degree distribution of a large graph generated by the static model has an analytical form as
\begin{eqnarray}
P(k) = \frac {1}{\xi}\frac {[c (1 - \xi)]^k}{ k!} {\rm E}_{- k + 1 + \frac{1}{\xi}} [c (1 - \xi)],
\end{eqnarray}
with $\xi \equiv 1/(\gamma - 1)$.
The special function ${\rm E}_{a}(x)$ is the general exponential integral function as
${\rm E}_{a}(x) \equiv \int _{1}^{\infty} \mathrm{d}t {\rm e}^{-xt} t^{- a}$ with $a, x > 0$.
For large $k$, we have $P (k) \propto k^{- \gamma}$.

For ensembles of random bipartite graphs, we consider the following types.
(1) Diluted RR bipartite graphs (RRRR):
the mean degree for out- (in-)vertices is $c_{\pm} = \rho K_{\pm}$,
in which $K_{+}$($K_{-}$) is the initial out-(in)-degree and $\rho \in [0, 1]$;
we consider $K_{+} = 24$ and $K_{-} \in \{48, 24, 20\}$.
(2) Bipartite graphs with both Poisson out- and in-degree distributions (ERER).
(3) Bipartite graphs with a Poisson out-degree distribution and a power-law in-degree distribution (ERSM):
the in-degree exponent is $\gamma _{-} = 3.0$.
(4) Bipartite graphs with a power-law out-degree distribution and a Poisson in-degree distribution (SMER):
the out-degree exponent is $\gamma _{+} = 3.0$.
(5) Bipartite graphs with both power-law out- and in-degree distributions (SMSM):
the out-degree exponent is $\gamma _{+} = 3.0$
and the in-degree exponent is $\gamma _{-} \in \{3.5, 3.0, 2.8\}$.
(6) Bipartite graphs with a Poisson out-degree distribution and a uniform in-degree distribution (ERRR):
the uniform in-degree is $k$.
(7) Bipartite graphs with a power-law out-degree distribution and a uniform in-degree distribution (SMRR):
the out-degree exponent is $\gamma _{+} \in \{3.5, 3.0, 2.8\}$
and the uniform in-degree is $k$.

In figures \ref{fig:rand_z1} and \ref{fig:rand_z2},
we show results of $z$-matchings and GLRB with $z = 1$ and $2$, respectively.
For $z$-matchings,
$y$ increases with $c_{+}$ from $0$ continuously and smoothly approaches its saturated size.
For GLRB, with small $c_{+}$,
there is a trivial core as $n_{+} = n_{-} = 0$ and an increasing $w$.
Thus, the local steps of GLRB remove all edges,
and the sizes of matched edges increase with iterative steps in GLRB on graphs.
At some critical value of $c_{+}$,
a macroscopic core emerges with $n_{+} > 0$ and $n_{-} > 0$,
and $w$ correspondingly undergoes a turn.
With larger $c_{+}$,
both $n_{+}$ and $n_{-}$ further increases and $w$ decreases,
since there is a smaller size of less connected vertices for local steps in GLRB.
In the case of $z = 1$ and $\phi = 1$
on random graphs with the same out- and in-degree distributions,
the emergence of core is continuous
\cite{Liu.Csoka.Zhou.Posfai-PRL-2012}.
Yet in most of the cases here, percolation phenomena are discontinuous,
except one of ERSM with $z = 1$ and $\phi = 0.5$.
We give here an explanative description for discontinuous core percolations below.

The discontinuity of percolation transitions is imbedded
in the behaviors of stable fixed solutions of $(\alpha ^{\pm}, \beta^{\mp})$.
We define the right-hand sides of equations (\ref{eq:alpha_p}) - (\ref{eq:beta_p}) as
$F_{+} (\beta ^{-})$, $G_{-} (\alpha ^{+})$, $F_{-} (\beta ^{+})$, and $G_{+} (\alpha ^{-})$, respectively.
Then we define $f_{\pm} (\alpha ^{\pm}) \equiv - \alpha ^{\pm} + F_{\pm} (G_{\mp} (\alpha ^{\pm}))$.
Finding the stable fixed solutions of equations (\ref{eq:alpha_p}) - (\ref{eq:beta_p})
is equivalent to finding the smallest fixed solutions of $\alpha ^{\pm}$ under $f_{\pm} (\alpha ^{\pm}) = 0$.
For equations (\ref{eq:alpha_p}) - (\ref{eq:beta_p}),
a substitution as $\alpha ^{\pm} \rightarrow 1 - \beta^{\pm}$ retains their forms.
Thus we have a trivial solution as $1 - \alpha ^{\pm} - \beta^{\pm} = 0$.
With small $c_{\pm}$, the trivial solution is stable,
and $n^{\pm} = 0$ based on equations (\ref{eq:n_p}) and (\ref{eq:n_m}).
With large enough $c_{\pm}$, the trivial solution becomes unstable,
and nontrivial stable solutions emerge as $1 - \alpha ^{\pm} - \beta^{\pm} > 0$,
leading to $n^{\pm} > 0$.
See figure \ref{fig:fixed} for an example on SMER with $\phi = 0.6$ and $z = 2$.
In figures \ref{fig:fixed} (a) and (b),
when $c_{+} < c_{+}^{\ast} \approx 3.9182$
with $c_{+}^{\ast}$ as the critical mean out-degree,
$f_{\pm}(\alpha ^{\pm}) = 0$
both has only one fixed, and also the stable, solution as $1 - \alpha ^{\pm} - \beta^{\pm} = 0$.
When $c_{+} > c_{+}^{\ast}$, new fixed solutions emerge.
There are different scenarios of behaviors for new stable solutions of $\alpha ^{+}$ and $\alpha ^{-}$.
For $\alpha ^{+}$,
the newly emerged two fixed solutions of $f_{+}(\alpha ^{+}) = 0$
are both larger than the previous one,
which means the branching of fixed solutions of $\alpha ^{+}$
brings only a continuous change to its stable fixed solution.
For $\alpha ^{-}$,
the two new fixed solutions of $f_{-}(\alpha ^{-}) = 0$
are both smaller than the previous one by a gap from $\approx 0.967$ to $\approx 0.052$,
leading to a sudden drop of its stable fixed solution
and further $1 - \alpha ^{\pm} - \beta^{\pm} > 0$.
More directly from figure \ref{fig:fixed} (c),
we can see that the stable fixed solutions of $(\alpha ^{+}, \beta ^{-})$ behavior continuously,
while those of $(\alpha ^{-}, \beta^{+})$ show a drop at $c_{+}^{\ast}$,
resulting in both a sudden birth of a core for $n^{\pm}$
and an abrupt shrinkage of sizes of matched edges for $w$.

\subsection{Comparison with Krea\v{c}i\'{c} and Bianconi's theory}

\begin{figure*}
\begin{center}
 \includegraphics[width = 0.99 \linewidth]{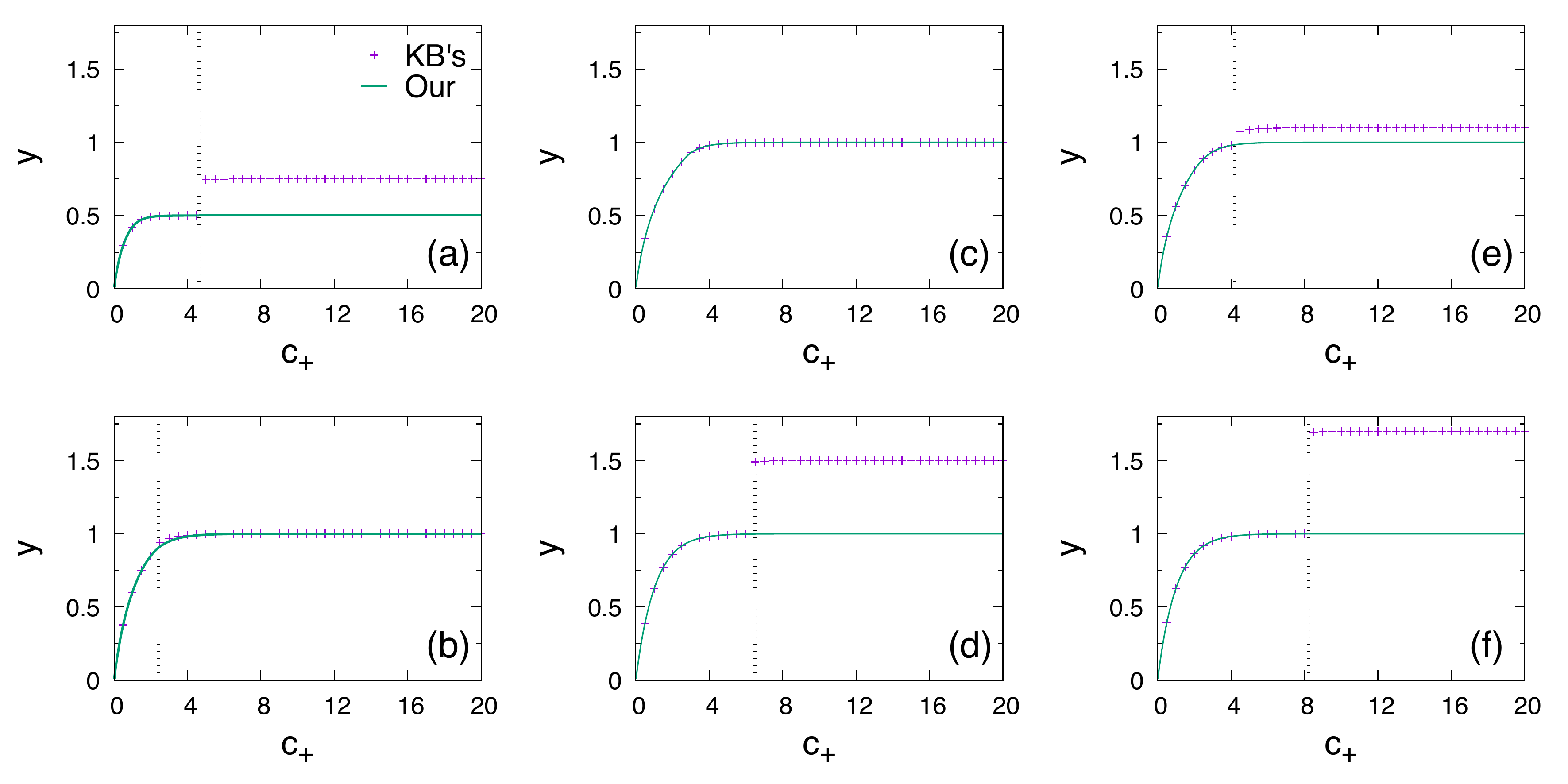}
\end{center}
\caption{
\label{fig:KB}
Fractions of $z$-matchings with the analytical theory in
\cite{Kreacic.Bianconi-EPL-2019}
and our theoretical framework on infinitely large ERER graphs.
(a), (b) Fractions of $z$-matching with $z = 1$ and $2$ respectively for $\phi = 0.5$.
(c), (d) and (e), (f) are for the cases similar to (a) and (b) yet with $\phi = 1.0$ and $1.2$, respectively.
Signs are for some trial results from the theory in \cite{Kreacic.Bianconi-EPL-2019}.
Solid lines are for results from our framework.
Dashed vertical lines denote critical $c_{+}$ for discontinuous percolation based on our framework.}
\end{figure*}
%

We also compare the belief propagation algorithm at zero-temperature limit from
\cite{Kreacic.Bianconi-EPL-2019}
with our theory on the $z$-matching fractions $y$.
We test them on ERER with different combinations of $\phi$ and $z$.
Results are shown in figure \ref{fig:KB}.
We can see that, 
our framework gives smooth and physically reasonable predictions,
no matter there is a continuous or discontinuous percolation.
The two frameworks generate the same results in two cases:
in the regime with no percolation, and in a simple case ($\phi = 1, z = 1$) with a continuous percolation in figure (\ref{fig:KB}) (c),
both of which are theoretically proved in the Theory section.
Yet when there is a discontinuous percolation, the analytical theory from
\cite{Kreacic.Bianconi-EPL-2019}
makes slightly higher predictions than ours around the percolation transition point in figure \ref{fig:KB} (b),
even unphysical ones as $y > 0.5$ in figure \ref{fig:KB} (a) and  $y > 1$ in figures \ref{fig:KB} (d)-(f).
Our guess is that in the theory of \cite{Kreacic.Bianconi-EPL-2019},
there are probably some errors in the derivation steps
related to contribution of matched edges from cores.

\subsection{Percolation analysis of $k$-XORSAT problem}

\begin{figure*}
\begin{center}
 \includegraphics[width = 0.99 \linewidth]{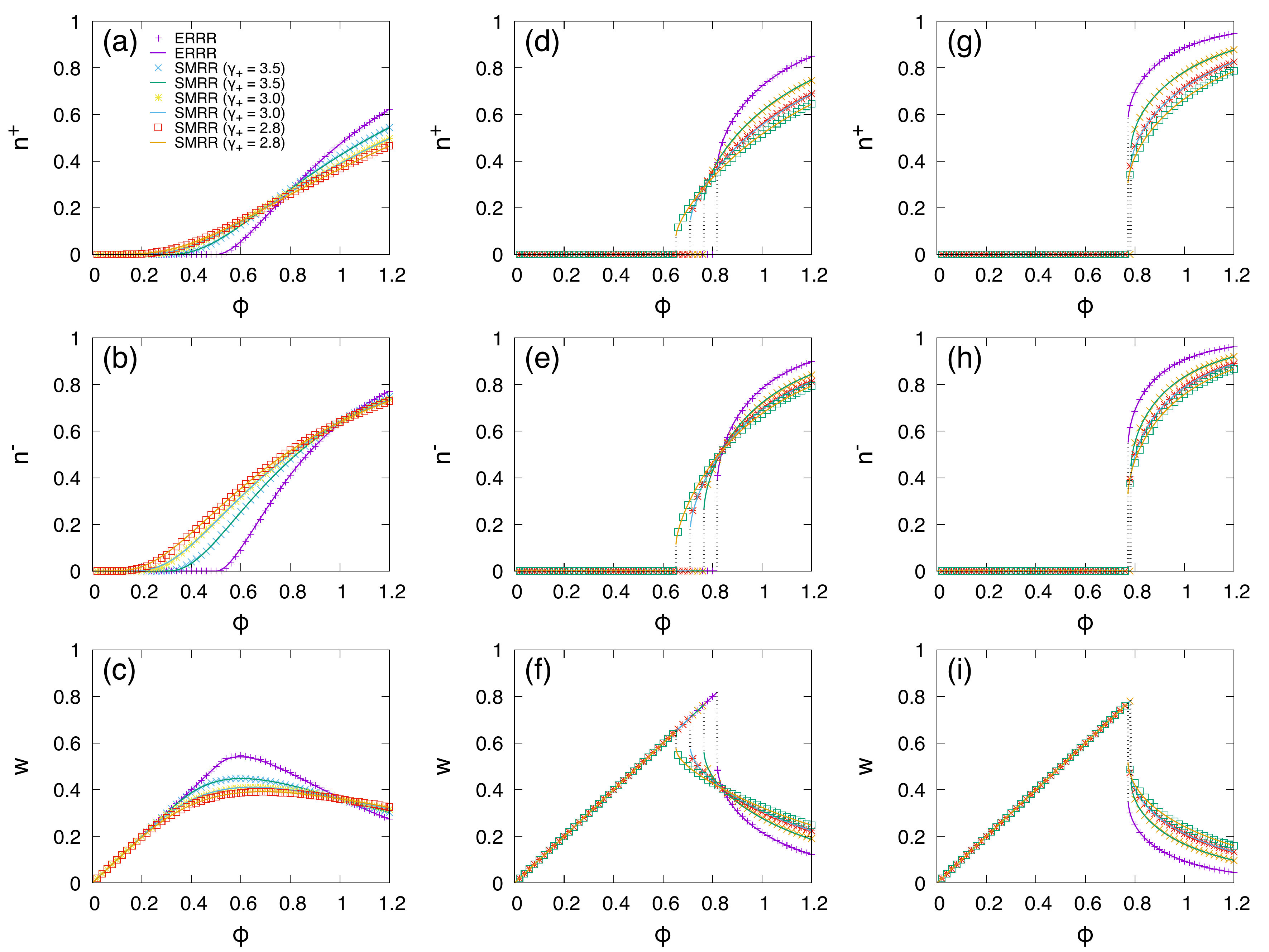}
\end{center}
\caption{
\label{fig:kxorsat}
Percolation analysis with our analytical framework for the leaf removal algorithm on hypergraphs
for the $k$-XORSAT problem.
Details of graph ensembles are in the main text.
(a)-(c) Results in the case of $k = 2$.
(d)-(f), (g)-(i) The similar case yet with $k = 3$ and $4$, respectively.
Signs are for simulation results on a graph instance with an out-vertex size $|V_{+}| = 10^5$.
Solids lines are for analytical results on infinitely large random graphs.
Dotted vertical lines denote discontinuous changes in analytical results.}
\end{figure*}

We test our analytical framework on the leaf removal algorithm for XORSAT problems on hypergraphs.
For simplicity, we only consider $k$-XORSAT problems with a uniform in-degree $k$ as $P_{-}(k_{-}) = \delta (k_{-}, k)$, yet with different out-degree distributions.
Results are shown in figure \ref{fig:kxorsat}.
We can see that simulation and analytical results correspond very well,
and discontinuous core percolations happen when $k \geqslant 3$.

In the specific case of $k = 3$, we further show that, the method based on dynamical evolution of degrees in graph pruning process in \cite{Cocco.etal-PRL-2003} and our percolation theory reach at the same iterative equations for a core percolation.
On graph ensembles with Poisson out-degree distributions,
combining iterative equations for $\alpha$ and $\beta$ as
equations (\ref{eq:xorsat_a}) and (\ref{eq:xorsat_b}),
we have
\begin{eqnarray}
1 - \alpha ^{+} = 1 - {\rm e}^{- \phi k (1 - \beta ^{-})} = 1 - {\rm e}^{- \phi k (1 - \alpha ^{+})^{k - 1}}.
\end{eqnarray}
Define $1 - \alpha ^{+} \equiv b$, $\phi \equiv c$, and set $k = 3$, we have
\begin{eqnarray}
b = 1 - {\rm e}^{- 3 c b^2}.
\end{eqnarray}
This is simply equation (2) in
\cite{Cocco.etal-PRL-2003} from a different derivation based on evolution of degree distributions in the leaf removal process.
As the entropy landscape in the clustering regime of $k$-XORSAT problem with $k \geqslant 3$
is described by $\alpha^{+}$ \cite{Cocco.etal-PRL-2003},
our framework thus provides a geometric perspective to its understanding.

\subsection{Result on a real-world network}

\begin{figure*}
\begin{center}
 \includegraphics[width = 0.75 \linewidth]{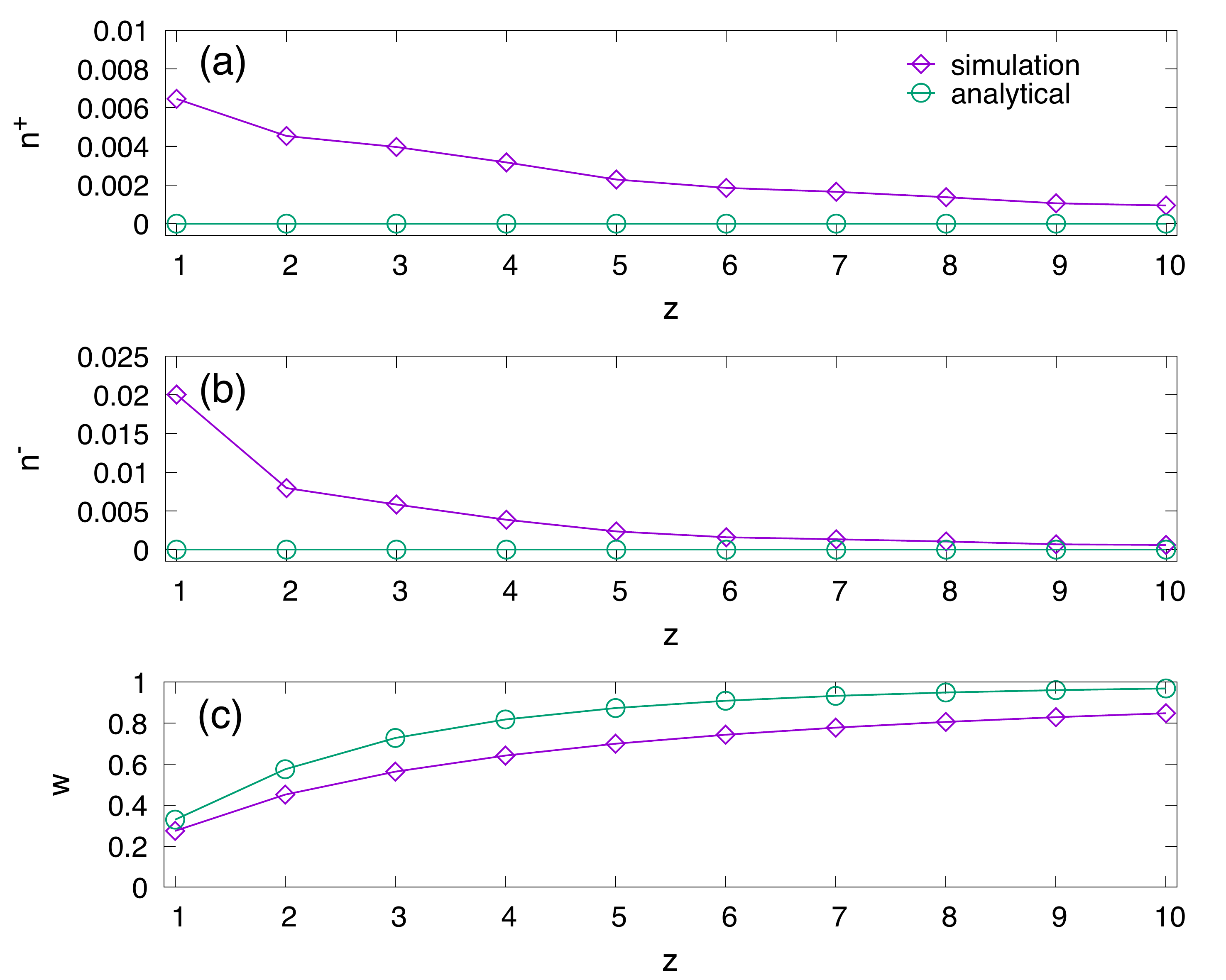}
\end{center}
\caption{
\label{fig:actor2}
GLRB on a movie-actor bipartite network. We consider cases of $z = 1$ to $10$.
Both simulation results and theoretical predictions of $n^{+}$, $n^{-}$, and $w$ are shown in (a) - (c), respectively.}
\end{figure*}

We further test our model and theory on real-world bipartite networks.
Specifically, we consider a movie-actor network extracted from IMDb\cite{Kunegis-2013}, which consists of $303,617$ in-vertices (movies), $896,302$ out-vertices (actors), and $3,782,463$ edges (an actor appearing in a movie).
Simulation results of GLRB on the network, including $n^{+}$, $n^{-}$, and $w$, can be found in figure \ref{fig:actor2}.
For comparison, we also show their theoretical predictions with the empirical degree distributions of out- and in-vertices as inputs into the analytical framework. Intrinsically, these theoretical predictions correspond to average simulation results on degree-preserving randomization of the network, in which connections are randomized while the degree of each vertex is kept unchanged.
We can see that the simulation results of core structures $n^{+}$ and $n^{-}$ are always larger than their theoretical predictions, while those sizes of matched edges $w$ are smaller.
The reason of this discrepancy is that,
our theoretical framework for GLRB is established on uncorrelated random graphs,
yet real-world networks show rich structure,
such as degree-degree correlation\cite{PastorSatorras.etal-PRL-2001},
communities\cite{Newman-NatPhys-2012},
spatial constraints\cite{Barthelemy-PhysRep-2011}, and so on.
All these factors can fail the mean-field approximation in our framework.
Yet consistently larger or smaller core structure on certain types of networks than their randomized instances could provide clues to their evolutional formation and topological organization. Characterizing and quantifying these clues can be a very interesting topic, and it can be examined when a large collection of datasets is available.

\section{Conclusion}
\label{sec:conclusion}

We study here the $z$-matching problem on bipartite graphs with a local algorithm.
The basic part of the local algorithm is a generalized GLR procedure extended to bipartite graphs,
in which four types of local graphical structures are identified and edges adjacent to them are selected and matched in local optimal steps.
We further develop a mean-field theory to quantitatively characterize this procedure on random bipartite graphs.
With an assumption of perfect matching of cores,
we reach at a concise analytical framework to directly estimate $z$-matching fractions
on random bipartite graphs with or without cores.
In all, our work here offers a simple example of combinatorial optimization problems on bipartite graphs, to which a direct analytical approach based on percolation theory can be taken.

Compared with the traditional matching problem on undirected and directed graphs,
the $z$-matching problem introduces two new dimensions to the problem:
the size ratio of two types of vertices $\phi$ and the parameter of matching size $z$.
For the theoretical framework here,
we can see that a nontrivial $z (\geqslant 2)$ makes the problem much more complicated to deal with.
Intuitively, our analytical framework operates at the replica symmetric level.
As far as we know,
their is no definite conclusion on the computational complexity type of the $z$-matching problem.
There are multiple related problems remained to answer,
such as whether the problem with $z \geqslant 2$ is still a polynomial problem or even a NP-hard problem,
and whether percolation transitions in our model exactly corresponds to clustering transitions
often shown in NP-hard problems.
To ascertain such issues, a detailed calculation at the replica symmetry breaking level should be carried out, which will be left in a future work.

\section*{Acknowledgements}

This work is supported by
Guangdong Major Project of Basic and Applied Basic Research No. 2020B0301030008 and
Guangdong Basic and Applied Basic Research Foundation (Grant No. 2022A1515011765).

\section*{Appendix A: GLRB as local optimal steps}

\begin{table}[htbp]
\caption{
 \label{tab:glrb_outleaf}
Comparison of matching $(i^{+}, j^{-})$ and $(m^{+}, j^{-})$
on $R(i^{+}, m^{+}, j^{-})$ and $M(i^{+}, m^{+}, j^{-})$.
In the units with two rows of values,
the upper and the lower one correspond to the case of $z_{j^{-}} > 0$ and $z_{j^{-}} = 0$, respectively.}
\begin{center}
\begin{tabular}{| c | c | c |}
\hline
 & $(i^{+}, j^{-})$ & $(m^{+}, j^{-})$ \\
 \hline
\multirow{2}*{$R_{(\cdot, \cdot)}(i^{+}, m^{+}, j^{-})$} & $1$ & $k_{m^{+}} $ \\
  & $k_{j^{-}}$ & $k_{j^{-}} + k_{m^{+}} - 1$ \\
\hline
\multirow{2}*{$(k_{i^{+}}, z_{i^{+}})$} & \multirow{2}*{$(0, 0)$} & $(1,1)$ \\
& & $(0,1)$ \\
\hline
\multirow{2}*{$(k_{m^{+}}, z_{m^{+}})$} & $(k_{m^{+}}, 1)$ & \multirow{2}*{$(k_{m^{+}} - 1,0)$} \\
& $(k_{m^{+}} - 1, 1)$ & \\
\hline
$(k_{j^{-}}, z_{j^{-}})$ & $(k_{j^{-}} - 1, z_{j^{-}} - 1)$ & $(k_{j^{-}} - 1, z_{j^{-}} - 1)$ \\
\hline
\multirow{2}*{$M_{(\cdot, \cdot)}(i^{+}, m^{+}, j^{-})$} & \multirow{2}*{$\min (k_{j^{-}}, z_{j^{-}})$} & $\min (k_{j^{-}}, z_{j^{-}})$ \\
& & $\min(k_{j^{-}} , z_{j^{-}}) - 1$ \\
\hline
\end{tabular}
\end{center}
\end{table}
\begin{table}[htbp]
\caption{
 \label{tab:glrb_inleaf}
Comparison of matching $(i^{+}, j^{-})$ and $(i^{+}, l^{-})$
on $R(i^{+}, j^{-}, l^{-})$ and $M(i^{+}, j^{-}, l^{-})$.}
\begin{center}
\begin{tabular}{| c | c | c |}
\hline
 & $(i^{+}, j^{-})$ & $(i^{+}, l^{-})$ \\
 \hline
$R_{(\cdot, \cdot)}(i^{+}, j^{-}, l^{-})$ & $k_{i^{+}}$ & $k_{i^{+}} $ \\
\hline
$(k_{i^{+}}, z_{i^{+}})$ & $(0, 0)$ & $(0,0)$ \\
\hline
$(k_{j^{-}}, z_{j^{-}})$ & $(k_{j^{-}}-1, z_{j^{-}} -1)$ & $(k_{j^{-}} - 1, z_{j^{-}})$ \\
\hline
$(k_{l^{-}}, z_{l^{-}})$ & $(k_{l^{-}} - 1, z_{l^{-}})$ & $(k_{l^{-}} - 1, z_{l^{-}} - 1)$ \\
\hline
$M_{(\cdot, \cdot)}(i^{+}, j^{-}, l^{-})$ & $k_{j^{-}} - 1 + z_{l^{-}}$ & $k_{j^{-}} - 1 + z_{l^{-}} - 1$ \\
\hline
\end{tabular}
\end{center}
\end{table}

We explain here how the two basic steps of GLRB works as local optimal methods
to construct a $z$-matching.

To compare different choices of matching edges in a bipartite graph,
we consider two local measures after matching an edge:
size of removed edges $R$, and maximal possible size of further matched edges $M$.
Specifically, we consider matching an edge $(m^{+}, n^{-})$
between an out-vertex $m^{+}$ and an in-vertex $n^{-}$
in a local structure consisting of $\partial i^{+}, \partial j^{-}, \partial \cdots$
for an out-vertex $i^{+}$, an in-vertex $j^{-}$, and so on.
Following GLRB, after matching $(m^{+}, n^{-})$,
$m^{+}$ is removed and $z_{n^{-}}$ is reduced by $1$.
If $z_{n^{-}} = 0$, $n^{-}$ will be further removed along with all its adjacent edges.
$R_{(m^{+}, n^{-})}(i^{+}, j^{-}, \cdots)$
is defined as the size of removed edges after matching $(m^{+}, n^{-})$
in a local structure with $\partial i^{+}, \partial j^{-}, \partial \cdots$.
$M_{(m^{+}, n^{-})}(i^{+}, j^{-}, \cdots)$
is defined as the maximal possible size of matching edges
after matching $(m^{+}, n^{-})$
in a local structure with $\partial i^{+}, \partial j^{-}, \partial \cdots$.
We have
$M_{(m^{+}, n^{-})}(i^{+}, j^{-}, \cdots)
= \sum _{i \in \{i^{+}, j^{-}, \cdots\}} M_{(m^{+}, n^{-})} (i)
= \sum _{i \in \{i^{+}, j^{-}, \cdots\}} \min (k_{i}, z_{i})$.
The two measures are compatible,
and the choice with smaller $R$ and larger $M$ is a better one.

First, we consider the step of matching an edge adjacent to an out-leaf.
As in figure \ref{fig:model} (b),
we compare two cases of matching $(i^{+}, j^{-})$ and $(m^{+}, j^{-})$,
in which $i^{+}$ is an out-leaf,
$j^{-}$ as its sole nearest neighbor,
and $m^{+}$ as a neighboring out-vertex of $j^{-}$ which is not an out-leaf.
We can see that $k_{m^{+}} \geqslant 2$.
$R(i^{+}, m^{+}, j^{-})$ and $M(i^{+}, m^{+}, j^{-})$
for two matching choices are listed in table \ref{tab:glrb_outleaf}.
We can see that
$R_{(i^{+}, j^{-})}(i^{+}, m^{+}, j^{-}) < R_{(m^{+}, j^{-})}(i^{+}, m^{+}, j^{-})$ and
$M_{(i^{+}, j^{-})}(i^{+}, m^{+}, j^{-}) \geqslant M_{(m^{+}, j^{-})}(i^{+}, m^{+}, j^{-})$.
Thus in a local sense,
matching $(i^{+}, j^{-})$ is the better choice for constructing a $z$-matching.

Then, we discuss the step of matching edges between an in-leaf and its out-roots.
As in figure \ref{fig:model} (c),
we compare the cases of matching $(i^{+}, j^{-})$ and $(i^{+}, l^{-})$,
in which $j^{-}$ is an in-leaf,
$i^{+}$ as a neighboring out-root of $j^{-}$,
and $l^{-}$ as a neighboring in-vertex of $i^{+}$.
We can see that $k_{j^{-}} \leqslant z_{j^{-}}$.
If $l^{-}$ is also an in-leaf,
matching $(i^{+}, j^{-})$ and $(i^{+}, l^{-})$ are locally equivalent.
To consider a non-trivial case, we take $l^{-}$ as not an in-leaf.
Then we have $k_{l^{-}} \geqslant z_{l^{-}} + 1$.
$R(i^{+}, j^{-}, l^{-})$ and $M(i^{+}, j^{-},l^{-})$
for two matching choices are listed in table \ref{tab:glrb_inleaf}.
We can see that
$R_{(i^{+}, j^{-})}(i^{+}, j^{-}, l^{-}) = R_{(i^{+}, l^{-})}(i^{+}, j^{-}, l^{-})$ yet
$M_{(i^{+}, j^{-})}(i^{+}, j^{-}, l^{-}) > M_{(i^{+}, l^{-})}(i^{+}, j^{-}, l^{-})$.
Thus in a local sense, matching $(i^{+}, j^{-})$ is the better choice.

\section*{Appendix B: Unique core structure on a graph after GLRB}

Here, we prove that on a given bipartite graph, the final core structure after GLRB is independent of the removal order of leaves and roots.

For a given graph $B = \{V_{+}, V_{-}, E\}$ upon GLRB, we assume that there are at least two different core configurations. For convenience, we consider here two configurations $S_{1}$ and $S_{2}$. Correspondingly, vertices in $B$ can be classified into four sets as $S_{\rm cc}$, $S_{\rm nn}$, $S_{\rm cn}$, $S_{\rm nc}$.
$S_{\rm cc}$ ($S_{\rm nn}$) denotes vertices both in (not in) the core structure in $S_{1}$ and $S_{2}$.
$S_{\rm cn}$ ($S_{\rm nc}$) denotes vertices in (not in) the core structure in $S_{1}$ but not in (in) the core structure of $S_{2}$.
These four sets form a mutually exclusive and collectively exhaustive division of the vertices in $B$.
To prove that the core structure is independent of the removal order is simply to prove that $S_{\rm cn} = S_{\rm nc} = \varnothing$.
 
If $S_{\rm cn} \neq \varnothing$, we focus on a vertex $i \in S_{\rm cn}$. 
As shown in the Model section, initially all vertices are considered as in a core structure until some or all of them are removed in an iterative process, in which a certain vertex among the four types of leaves and roots is removed deterministically and irreversibly along with its edges as in definition.
For the vertex $i$, the state change from being in a core structure in $S_{1}$ to not being in a core structure in $S_{2}$ is induced by some of its nearest neighbors, which are also in $S_{\rm cn}$.
The reason is that, those nearest neighbors of $i$ in $S_{\rm cc}$ and $S_{\rm nn}$ have the same states  (in a core or not in a  core) in both $S_{1}$ and $S_{2}$, thus they cannot contribute to the different states of $i$ in $S_{1}$ and $S_{2}$.
Furthermore, the vertices in $S_{\rm nc}$ which are not in a core in $S_{1}$ and are in a core in $S_{2}$ can only increase the chance of its nearest neighbors to be in a core in $S_{2}$, rather than decrease it. Thus those nearest neighbors of $i$ in $S_{\rm nc}$, if there are, do not contribute to $i$'s state change from $S_{1}$ to $S_{2}$.

For the vertex $i \in S_{\rm cn}$, we denote a set $S_{i}^{(1)}$ for the immediate vertices whose state change from being in a core to not being in a core directly induces $i$ also from being in a core to exit the core in $S_{2}$. With the same above procedure, we can find $S_{i}^{(2)}$ for $S_{i}^{(1)}$, $S_{i}^{(3)}$ for $S_{i}^{(2)}$, and so on. Since $S_{\rm cn}$ is a finite set, we can find the final $S_{i}^{(m)}$, with an integer $m \geqslant 1$. Yet for the vertices in $S_{i}^{(m)}$, neither of their nearest neighbors in $S_{\rm cc}$, $S_{\rm nn}$, and $S_{\rm nc}$ can induce a state change from being in a core in $S_{1}$ to not being in a core in $S_{2}$, thus vertices in $S_{i}^{(m)}$ can only have the same state in both $S_{1}$ and $S_{2}$, or equivalently being in $S_{\rm cc}$ or $S_{\rm nn}$. Thus a contradiction happens here. Thus we prove that $S_{\rm cn} = \varnothing$.

For two core configurations $S_{1}$ and $S_{2}$ here, since there is no intrinsically order among them, we can easily switch their order indices. With a similar logic, we can prove that $S_{\rm nc} = \varnothing$.
Finally, we have $S_{\rm cn} = S_{\rm nc} = \varnothing$.

\section*{Appendix C: Modified equations in analytical framework}

Equations of the analytical framework in the main text can be modified as
\begin{eqnarray}
\beta ^{-}
&& = 1 - \sum _{s = 0}^{z-1} (\alpha ^{+})^{s}
\times
\sum _{k_{-} = s+1}^{+\infty}  Q_{-}(k_{-})
{k_{-} - 1 \choose s}
(1 - \alpha ^{+})^{k_{-} -1 - s},\\
\alpha ^{-}
&& = \sum _{s = 0}^{z - 1}
(1 - \beta^{+})^{s}
\times
\sum _{k_{-} = s+1}^{+\infty}
Q_{-}(k_{-})
{k_{-} - 1 \choose s}
(\beta^{+})^{k_{-} - 1 -s},\\
n^{+}
&& = \sum _{k_{+} = 0}^{+ \infty} P_{+}(k_{+})
(1 - \alpha ^{-})^{k_{+}}
- \sum _{k_{+} = 0}^{+ \infty} P_{+}(k_{+})
(\beta ^{-})^{k_{+}}
- c_{+} \alpha ^{+} (1 - \alpha ^{-} - \beta^{-}), \\
n^{-}
&& =  \sum _{t = 0}^{z - 1}
(\alpha ^{+})^{t}
[\sum _{k_{-} = t}^{+\infty}
P_{-}(k_{-})
{k_{-} \choose t}
 (1 - \alpha ^{+})^{k_{-} - t} \nonumber \\
 &&
- \sum _{s = 0}^{z - t} (1 - \alpha ^{+} - \beta ^{+})^{s}
{t + s \choose s}
\times
\sum _{k_{-} = t + s}^{+\infty}
P_{-}(k_{-}) {k_{-} \choose t + s} (\beta ^{+})^{k_{-} - t - s}], \\
w
&& = [1 - \sum _{k_{+} = 0}^{+\infty}
P_{+}(k_{+})(1 - \alpha ^{-})^{k_{+}}]
+ \phi z [1
- \sum _{s = 0}^{z - 1} (\alpha ^{+})^{s} \times
\sum _{k_{-} = s}^{+ \infty}
P_{-}(k_{-}) {k_{-} \choose s} (1 - \alpha ^{+})^{k_{-} - s}] \nonumber \\
&&
+ \phi \sum _{t = 1}^{z - 1}
t  (\alpha ^{+})^{t}
[\sum _{k_{-} = t}^{+\infty}
P_{-}(k_{-})
{k_{-} \choose t}
 (1 - \alpha ^{+})^{k_{-} - t} \nonumber \\
&&
- \sum _{s = 0}^{z - t} (1 - \alpha ^{+} - \beta ^{+})^{s}
{t + s \choose s}
\times
\sum _{k_{-} = t + s}^{+\infty}
P_{-}(k_{-}) {k_{-} \choose t + s} (\beta ^{+})^{k_{-} - t - s}] \nonumber \\
&&
- \phi z (\alpha ^{+})^{z} \times
\sum _{k_{-} = z}^{+ \infty}
P_{-}(k_{-}) {k_{-} \choose z} (\beta ^{+})^{k_{-} - z}, \\
n_{\rm M}^{-}
&& =  \sum _{t = 0}^{z - 1} (z - t)
(\alpha ^{+})^{t}
[\sum _{k_{-} = t}^{+\infty}
P_{-}(k_{-})
{k_{-} \choose t}
 (1 - \alpha ^{+})^{k_{-} - t} \nonumber \\
 &&
- \sum _{s = 0}^{z - t} (1 - \alpha ^{+} - \beta ^{+})^{s}
{t + s \choose s}
\times
\sum _{k_{-} = t + s}^{+\infty}
P_{-}(k_{-}) {k_{-} \choose t + s} (\beta ^{+})^{k_{-} - t - s}].
\end{eqnarray}

The sums of $\sum _{k = s + 1}^{+\infty} Q(k) {k - 1 \choose s} x^{k - 1 - s}$ and
$\sum _{k = s}^{+\infty} P(k) {k \choose s} x^{k - s}$,
with a parameter $x \in [0, 1]$ and an integer $s \geqslant 0$, are also the generating functions of $Q(k)$ and $P(k)$ respectively. For random graphs with analytical degree distributions, both two summations can be put in a compact form. Inserting these summation results into the above modified equations, we can easily numerically solve our analytical framework on random graph ensembles.

\end{document}